# Third-order nonlinear optical response of 2D materials in the telecom band

**David J. Moss**

Optical Sciences Center, Swinburne University of Technology, Hawthorn, VIC 3122, Australia;
dmoss@swin.edu.au (D. M.).

**Abstract:** All-optical signal processing based on nonlinear optical devices is promising for ultrafast information processing in optical communication systems. Recent advances in two-dimensional (2D) layered materials with unique structures and distinctive properties have opened up new avenues for nonlinear optics and the fabrication of related devices with high performance. This paper reviews the recent advances in research on third-order optical nonlinearities of 2D materials, focusing on all-optical processing applications in the optical telecommunications band near 1550 nm. First, we provide an overview of the material properties of different 2D materials. Next, we review different methods for characterizing the third-order optical nonlinearities of 2D materials, including the Z-scan technique, third-harmonic generation (THG) measurement, and hybrid device characterization, together with a summary of the measured $n_2$ values in the telecommunications band. Finally, the current challenges and future perspectives are discussed.

## 1. Introduction

All-optical signal processing based on nonlinear optical devices is an attractive technique for ultrahigh speed signal processing for optical communication systems. It offers broad operation bandwidths, ultra-high processing speeds, together with low power consumption and potentially reduced footprint and cost. Integrated nonlinear optical photonic chips have been based on a few key materials including silicon (Si) [1-3], doped silica ($SiO_2$) [4, 5], silicon nitride ($Si_3N_4$) [6, 7], aluminum gallium arsenide (AlGaAs) [8-10], and chalcogenide glasses [11, 12]. These have enabled a wide range of devices from Raman amplification and lasing [13-15], wavelength conversion [5, 12, 16-18], optical logic gates [19-22], and optical frequency comb generation [23-26], to optical temporal cloaking [27], quantum entangling [28-30], and many others. Despite their success, no platform is perfect – they all have limitations, such as a relatively small Kerr nonlinearity ($n_2$) (e.g., for $Si_3N_4$) or high two photon absorption (for silicon in the telecommunications band), resulting in a low nonlinear figure of merit (FOM = $n_2 / (\lambda \beta_{TPA})$, with $n_2$ and $\beta_{TPA}$ denoting the effective Kerr coefficient and TPA coefficient of the waveguides, respectively, and $\lambda$ the light wavelength).

To overcome these limitations, newly emerging materials have attracted significant attention, particularly 2D layered materials, such as graphene [31-33], GO [34-36], TMDCs [37-40], h-BN [41-43], and BP [44-46], where their atomically thin nature yields unique and superior optical properties. In particular, their properties are highly dependent on the number of atomic layers – not only is their optical bandgap highly layer thickness dependent but they can also exhibit an indirect-to-direct bandgap transition (and the reverse), which provides powerful ways in which to tune their optical responses [37, 46-48]. Further, their broadband photoluminescence and ultrahigh carrier mobility are highly attractive features for photonic and optoelectronic applications [33, 49-53]. Finally, in addition to their linear optical properties, 2D materials exhibit remarkable nonlinear optical properties including strong saturable absorption (SA) [54-57], a giant Kerr nonlinearity [58-62], and prominent second- (SHG) and third-harmonic generation (THG) [44, 63-65], opening up new avenues for high-performance nonlinear optical devices.

In contrast to the second-order optical nonlinearity that only exists in non-centrosymmetric materials, the third-order susceptibility is present in all materials, which gives rise to a rich variety of processes, including four-wave mixing (FWM), self-phase modulation (SPM), cross-phase modulation (XPM), THG, two-photon absorption (TPA), SA, stimulated Raman scattering, and many others. These third-order nonlinear optical processes are quasi-instantaneous with ultrafast response times on the order of femtoseconds [66]. This has motivated ultrafast all-optical signal generation and processing for telecommunications, spectroscopy, metrology, sensing, quantum optics, and many other areas [67, 68].

In this paper, we review recent progress in the study of the third-order optical nonlinearities of 2D materials specifically in the telecommunications wavelength band near 1550 nm, in contrast with other reviews [44, 69, 70] that focus predominantly in the visible wavelength range. We discuss the different techniques for characterizing the third-order optical nonlinearity and the prospects for future development. In Section 2, the material properties of different 2D materials are briefly introduced and compared. Next, we review different methods used to characterize the third-order nonlinear optical response of 2D materials, including Z-scan technique, THG measurement, and hybrid device characterization. We also summarize the measured values of $n_2$ of different 2D materials in the telecommunications band. Finally, the conclusion and future perspectives are discussed in Section 4.

## 2. 2D materials

The past decade has witnessed an enormous surge in research on layered 2D materials – many have been discovered and synthesized with a wide range of properties. In this section, we briefly introduce some key 2D materials such as graphene, GO, TMDCs, h-BN, and BP and discuss their electrical and optical properties.

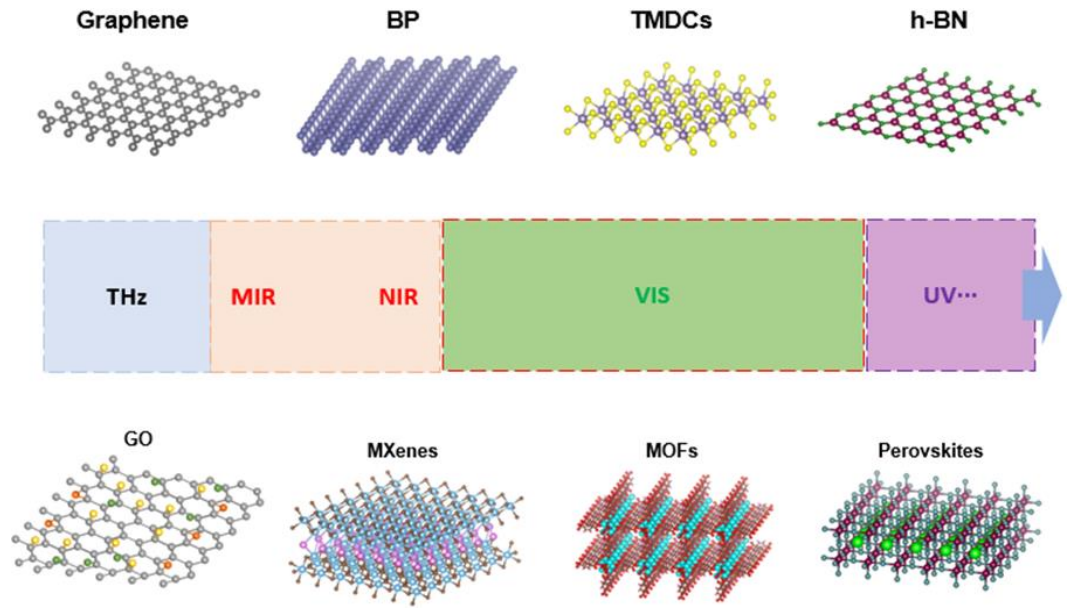

**Figure 1.** Illustration of typical 2D layered materials.

2.1 Graphene and graphene oxide

Graphene, and its derivative, graphene oxide (GO), have been intensely studied due to their excellent mechanical, electrical, and optical properties [33, 71, 72]. Graphene has a gapless band structure, in which the conduction and valence bands meet at the K point of Brillouin zone, resulting in its semimetal nature [31, 44, 73]. In contrast, GO is an electronically hybrid material, featuring both conducting $\pi$-states from $sp^2$ carbon sites and a large energy gap between the $\sigma$-states of its $sp^3$-bonded carbons [34, 74]. Their unique band structures result in novel electrical and optical properties, where for graphene, for example, the electrons and holes act as massless Dirac fermions resulting in extremely high carrier mobilities ( >$10^5$ $cm^2$/Vs) even under ambient conditions [31]. In contrast, GO exhibits a band gap that is tunable by adjusting the degree of reduction, which in turn affects the electric and optical properties. In addition, GO exhibits fluorescence in the near-infrared (NIR), visible and ultraviolet regions [34-36], which is very promising for light emitting devices. Moreover, the excellent nonlinear optical properties of both materials have been reported, including strong saturable absorption (SA) [75, 76], a giant optical Kerr nonlinearity [58, 59], leading to efficient self-phase modulation [77], FWM [78, 79], as well as high harmonic generation [63].

2.2 Transition metal dichalcogenides

Transition metal dichalcogenides (TMDCs) with the formula of $MX_2$ (where M is a transition metal and X is a chalcogen), is another widely studied family of 2D materials. Different to the semimetal graphene, monolayer TMDCs, such as $MoS_2$, $MoSe_2$, $WS_2$, and $WSe_2$, are typically semiconductors that have bandgaps from 1 eV to 2.5 eV, covering the spectral range from the near infrared to the visible region [37, 38]. Moreover, TMDCs can exhibit a transition from direct- to indirect-bandgaps with increasing film thickness, resulting in strongly thickness-tunable optical and electrical properties. For instance, $MoS_2$ exhibits layer-dependent photoluminescence, with monolayer films showing a much stronger photoluminescence [80]. Monolayer hexagonal TMDCs also exhibit unique band structure valley-dependent properties, such as valley coherence and valley-selective circular dichroism [37, 81], offering new prospects for novel applications in optical computing and information processing. For the nonlinear optical properties, TMDCs with odd numbers of layers have no inversion symmetry, and so exhibit a non-zero second-order (and higher even-order) nonlinearities that are absent in graphene and even-layer TMDCs [44, 64]. Recently, noble metal TMDCs, including $PdSe_2$ and $PtSe_2$, and $PdTe_2$, have also attracted increasing interest in the fabrication of high performance electronic and optical devices, such as ultra-broadband photodetectors [39, 40] as well as mode-locked lasers [82].

2.3 Black phosphorus

Black phosphorus (BP) is another attractive single element 2D layered material which has been widely studied. It has a puckered crystal structure, yielding a strong in-plane anisotropy in its physical properties in the "armchair" and "zigzag" directions, opening new avenues for anisotropic electronic and optoelectronic devices [44, 45, 83]. Moreover, BP is a semiconductor that features a layer thickness dependent direct bandgap from 0.3 eV (bulk) to 2.0 eV (monolayer), bridging the gap between the zero-bandgap graphene and large-bandgap TMDCs [48, 83]. This broad bandgap tunability is very suitable for the photodetection and photonic applications from the visible to mid-infrared spectra regions [46, 47, 84]. For the nonlinear optical properties, the layer thickness tunable and polarization dependent THG and optical Kerr nonlinearity have been demonstrated recently [65, 83, 85]. Broadband SA has also been observed in BP, demonstrating its strong potential for ultrafast pulsed lasers [86-88].

2.4 Other emerging 2D materials

A wide range of other novel 2D low-dimensional materials have been investigated, including h-BN, MXenes, perovskites, as well as MOFs, which greatly enriches the family of 2D materials. h-BN is an electrical insulator with a large bandgap of around 5.9 eV [41, 89] making h-BN a candidate for ultraviolet light applications. It also has an ultra-flat surface as well as excellent resistance to oxidation and corrosion, which are both highly useful as a dielectric or capping layer to protect the active materials or devices from degradation [41].

MXenes belong to another family of 2D materials, including 2D transition metal carbides, nitrides, and carbonitrides. Typically, the electronic structure of MXenes can be tuned by varying the surface functional groups. For instance, nonterminated $Ti_3C_2$ theoretically resembles a typical semimetal with a finite density of states at the Fermi level, whereas it can transition to a semiconductor when terminated with surface groups, such as OH and F groups [90]. MXnens also exhibit superior optical properties, such as a high optical transmittance of visible light ( > 97% per nm) [38, 91], and excellent nonlinear optical properties [57].

Organometal-halide perovskites have a general formula of $ABX_3$, where typically A = $CH_3NH_3^+$, B = $Pb^{2+}$, and X = $I^-$, $Br^-$, $Cl^-$ or mixtures [92]. Due to their prominent photovoltaic features and luminescence properties, organometal-halide perovskite semiconductors have been widely used to design high performance solar cells as well as light-emitting diodes [51-53]. Metal-organic frameworks (MOFs) are organic–inorganic hybrid porous crystalline materials with metal ions or metal-oxo clusters coordinated with organic linkers [93, 94]. Thanks to this unique structure, 2D MOFs exhibit enhanced photo-physical behaviour and are promising for various applications, from light emission and sensing to nonlinear optical applications [95, 96].

**3. Third-order optical nonlinearities of 2D materials in the telecommunications band**

With their excellent third-order optical nonlinearities, 2D materials are promising functional materials for high-performance nonlinear optical devices. In this section, we review the different methods used to characterize their third-order nonlinear optical response. These include the Z-scan technique, THG measurement, and hybrid device characterization. We also summarize and compare the measured $n_2$ values of different 2D materials in the telecommunications band.

3.1 Third-order optical nonlinearity

The nonlinear optical response of a material in the dipole approximation is given by [1, 97]:

$$\tilde{P}(t) = \varepsilon_0[\chi^{(1)} \cdot \tilde{E}(t) + \chi^{(2)} : \tilde{E}(t)\tilde{E}(t) + \chi^{(3)} \vdots \tilde{E}(t)\tilde{E}(t)\tilde{E}(t) + \cdots ] \quad (1)$$

where the $\tilde{P}(t)$ is the material electronic polarization, $\tilde{E}(t)$ is the incident field, $\chi^{(n)}$ are the $n^{th}$-order nonlinear optical susceptibility. The first-order term $\chi^{(1)}$ describes the linear refractive index including refraction and absorption and is a result of the dipole response of bound and free electrons to a single photon [1]. The second-order term $\chi^{(2)}$ is a 3rd rank tensor, nonzero only for non-centrosymmetric materials, describes second-harmonic generation (SHG), sum- and difference frequency generation (SFG, DFG), optical rectification, the Pockels effect and others. The third-order nonlinear optical susceptibility $\chi^{(3)}$ is particularly important because it exists in all materials regardless of the crystal symmetry and gives rise to a rich variety of nonlinear processes, represented by THG [1, 97], FWM [78, 98], SPM [61, 99], and XPM [100, 101]. These form the basis of all-optical processing devices, such as wavelength conversion, optical comb generation, quantum entanglement, and more.

Equation (2) gives a simple description of the relevant third-order nonlinear optical effects corresponding to $\tilde{P}^{(3)}(t) = \varepsilon_0 \chi^{(3)} \cdot \tilde{E}^3(t)$ [97] as follows:

$$\tilde{P}^{(3)}(t) = \varepsilon_0 \int_{-\infty}^{\infty} \frac{d\omega_1}{2\pi} \int_{-\infty}^{\infty} \frac{d\omega_2}{2\pi} \int_{-\infty}^{\infty} \frac{d\omega_3}{2\pi} \chi^{(3)}(\omega_\sigma; \omega_1, \omega_2, \omega_3) \times E(\omega_1)E(\omega_2)E(\omega_3)e^{-i\omega_\sigma t} \quad (2)$$

where $\omega_\sigma = \omega_1 + \omega_2 + \omega_3$, with $\omega_1$, $\omega_2$, and $\omega_3$ denoting the angular frequencies. Different $\chi^{(3)}$ effects can be described with different wave frequency combinations, such as THG ( $\chi^{(3)}(\omega_\sigma = 3\omega_1; \omega_1, \omega_1, \omega_1)$ ), non-degenerate FWM ( $\chi^{(3)}(\omega_\sigma = \omega_1 + \omega_2 - \omega_3; \omega_1, \omega_2, -\omega_3)$) and degenerate FWM ($\chi^{(3)}(\omega_\sigma = 2\omega_1 - \omega_2; \omega_1, \omega_1, -\omega_2)$).

A key component of $\chi^{(3)}$ given by $n_2 = 3 \cdot \mathrm{Re}[\chi^{(3)}/4cn_0^2\varepsilon_0]$, where $n_2$ is the intensity-dependent refractive index change, known as the Kerr effect and the complex refractive index $n$ can be expressed as [1, 97]:

$$n = n_0 + n_2 I - i\frac{\lambda}{4\pi}(\alpha_0 + \alpha_2 I) \qquad (3)$$

where $I$ is the light intensity, $\lambda$ is the wavelength, $n_2$ represents the Kerr coefficient or Kerr nonlinearity, $\alpha_2$ is the nonlinear absorption induced by the third-order susceptibility $\chi^{(3)}$, and $n_0$, $\alpha_0$ are the linear refractive index and absorption, respectively.

In this paper, we focus on $\chi^{(3)}$ of 2D materials for key nonlinear processes that form the basis for ultra-high speed all-optical signal generation and processing, with response times on the order of femtoseconds [102, 103]. These include SPM and XPM, governed largely by $n_2$ via the Re ($\chi^{(3)}$), as well as FWM and THG that are mainly governed by the magnitude of $|\chi^{(3)}|$, although the latter are also sensitive to the complex value of $\chi^{(3)}$ via phase-matching effects. The $n_2$ component of $\chi^{(3)}$, accounts for two-photon absorption (TPA) via the Im ($\chi^{(3)}$) and can also result in saturable absorption (SA). Both are intrinsic functions of the material's bandgap, but can also be influenced by free carrier effects. At photon energies well below the bandgap, all $\chi^{(3)}$ components will become degenerate, but near, or above, the bandgap, they will in general be quite different. Finally, since nonlinear absorption is always present, it will affect the efficiency of all 3rd order nonlinear optical processes, not just n2, even though it does not arise directly from other $\chi^{(3)}$ components such as THG and FWM, for example. Further, these processes will generally scale differently with pump power to n2, and so the conventional nonlinear FOM may not be a useful benchmark.

3.2 Characterization methods

*3.2.1 Z-scan technique*

Measuring the Kerr coefficient of a material is needed in order to design and fabricate nonlinear optical devices. The Z-scan method, introduced in the 1990s [104] is an elegant method to measure the third-order optical Kerr nonlinearity of a material. This technique involves open-aperture (OA) and closed-aperture (CA) measurements, which can be used to measure the third-order nonlinear absorption and nonlinear refraction, respectively. CA Z-scan method is widely used to measure the nonlinear refractive index (Kerr coefficient) of an optical material. The valley-peak and peak-valley transmission curves are the typical results of the CA measurement, as shown in Figure 2(a). When the nonlinear material has a positive nonlinear refractive index ($n_2$ > 0), self-focusing will occur which results in the valley-peak transmission curve. The peak-valley CA curve arises from de-focusing and occurs with a negative nonlinear refractive index ($n_2$ < 0).

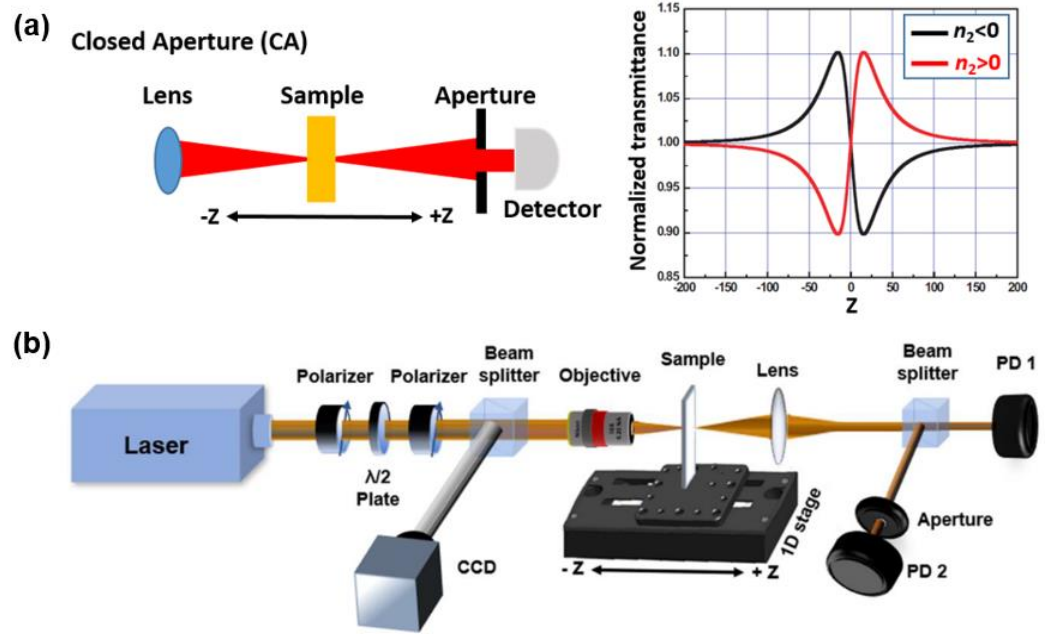

**Figure 2**. (a) Schemes showing the principle of closed-aperture (CA) Z-scan. (b) A typical Z-scan setup: PD: power detector, CCD: charge-coupled-device [62].

Figure 2(b) shows a typical Z-scan setup [62]. To measure the ultrafast nonlinear response, a femtosecond pulsed laser is used to excite the samples. A half-wave plate combined with a linear polarizer can be employed to control the power of the incident light. The beam is focused onto the sample with a lens or an objective. During the measurements, samples are oriented perpendicular to the beam axis and translated along the Z axis with a linear motorized stage. For the measurements of small micrometer sized samples, a high-definition charge-coupled-device imaging system can be employed to align the light beam to the target area. Two PDs are employed to detect the transmitted light power for the signal and reference arms.

For the CA Z-scan method, the normalized transmittance can be written as [62, 104]:

$$T(z, \Delta\Phi_0) \simeq 1 + \frac{4\Delta\Phi_0 x}{(x^2+9)(x^2+1)} \quad (4)$$

where $x = z/z_0$, $z_0 = k\omega_0^2/2$ with $\omega_0$ the beam waist radius and $k$ the wave vector. $\Delta\Phi_0$ represents the on-axis phase shift at the focus, is defined as [62, 104]:

$$\Delta\Phi_0 = k n_2 I_0 L_{eff} \quad (5)$$

In equation (5), $L_{eff} = (1 - e^{-\alpha L})/\alpha$, with $L$ denoting the sample length and $\alpha_0$ the linear absorption coefficient, $k$ is the wave vector which is defined by $k = 2\pi/\lambda$, and $I_0$ is the laser irradiance intensity with in the sample [104]. Based on the measured Z-scan curves, one can derive the Kerr coefficient $n_2$ with the fitting equations.

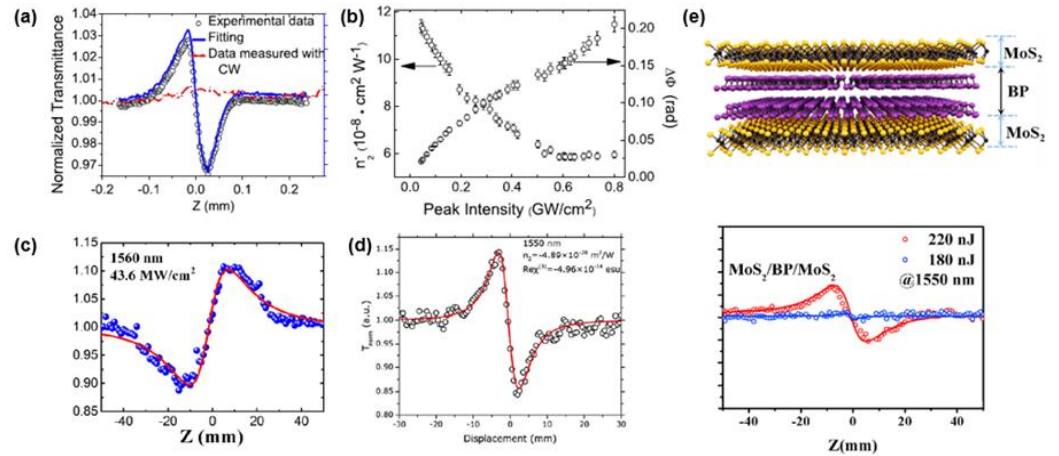

**Figure 3**. (a) CA Z-scan result of graphene under an excitation laser wavelength at 1550 nm. (b) Measured $n_2$ of graphene as a function of laser intensity [105]. (c) CA Z-scan result of $CH_3NH_3PbI_3$ under an excitation laser wavelength at 1560 nm [106]. (d) CA Z-scan result of MXene films under an excitation laser wavelength at 1550 nm [57]. (e) CA Z-scan result of $MoS_2/BP/MoS_2$ heterostructure at different laser intensities. The excitation laser wavelength is 1550 nm [107].

Graphene is the first 2D material to have been discovered, and its optical nonlinearities have been widely studied using Z-scan measurements and other methods. Figure 3(a) shows the CA Z-scan signal of a graphene film with an excitation laser wavelength at 1550 nm.[105] A peak-valley configuration can be observed, indicating a negative Kerr nonlinearity. The measured Kerr coefficient $n_2$ of graphene is as large as $10^{-11}$ m$^2$/W which is about 6 orders of magnitude larger than bulk Si, demonstrating the strong potential of 2D materials for nonlinear optical devices. A laser peak intensity dependent $n_2$ has also been observed (Figure 3(b)), providing a potential method for modulating its nonlinear properties. Figures 3(c) and (d) show the CA curves of $CH_3NH_3PbI_3$ perovskite[106] and $Ti_3C_2T_x$ MXene films [57] measured at a wavelength of 1550 nm, where a positive and negative Kerr nonlinearity were observed, respectively. The different response of these two materials forms the basis of their applications in different functional devices. For example, a negative Kerr nonlinearity can be used to self-compress ultrashort pulses in the presence of positive group-velocity dispersion while the materials with positive nonlinearity are promising for achieving a net parametric modulational instability gain under abnormal dispersion conditions.

2D van der Waals (vdW) heterostructures offers many new features and possibilities beyond what a single material can provide, and there has been significant activity in this field [108, 109]. Recently, the optical nonlinear response of 2D heterostructures has also been investigated via the Z-scan method. Figure 3(e) plots the CA curve of a $MoS_2/BP/MoS_2$ heterostructure at different laser intensities [107]. A negative Kerr nonlinearity at the telecommunications wavelength of 1550 nm can be observed. The strong Kerr nonlinearity of graphene/$Bi_2Te_3$ at the same wavelength was also demonstrated recently [110]. By fitting the experimental data, a large $n_2$ of $\sim 2 \times 10^{-12}$ m$^2$/W was obtained, which is highly attractive for all-optical modulators and switches.

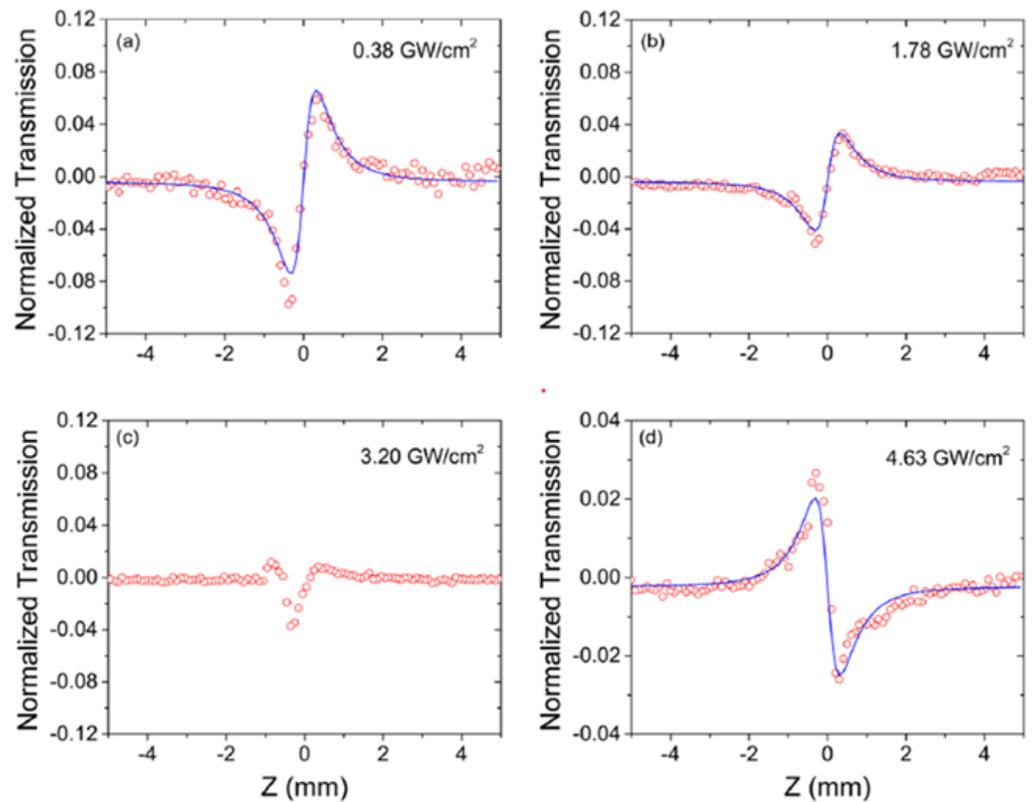

**Figure 4**. The CA Z-scan results of GO films under different irradiances: (a) 0.38 GW/cm$^2$; (b) 1.78 GW/cm$^2$; (c) 3.20 GW/cm$^2$; (d) 4.68 GW/cm$^2$ [60].

One of the unique features of GO is its tunable optical and electrical properties through laser reduction, which is particularly attractive for nonlinear optical applications. To investigate laser tunable optical nonlinearities, an in-situ third-order Kerr nonlinearity measurment for GO films has been conducted with the Z-scan method [60]. Figures 4(a)-(d) show the CA signal of GO films at different laser intensities. At low intensity, GO exhibits a positive Kerr nonlinearity with a valley-peak CA configuration. With increasing the laser intensity, GO reduction occurs and the positive nonlinearity finally transitions into a negative nonlinearity at an intensity of 4.63 GW/cm$^2$, at which point GO completely reduces to graphene. In addition to the ability to laser tune optical nonlinearities in GO, the measured Kerr coefficient $n_2$ of GO is as large as 4.5 × 10$^{-14}$ m$^2$/W at 1550 nm, which is four orders of magnitude higher than single crystalline silicon. These properties render GO a promising candidate for nonlinear applications in the telecommunications band.

*3.2.2 THG measurement*

In addition to Z-scan method, another technique that can be used to directly characterize the third-order optical nonlinearity of a material is THG measurement. As introduced in section 3.1, THG is a fundamental third-order optical nonlinear process in which three photons at the same frequency ($\omega_1$) excite the nonlinear media to generate new signal ($\omega = 3\omega_1$). Measuring the THG of a material provides a direct method to characterize its third-order optical nonlinearity. Figure 5 shows a typical setup for THG measurements[111] where a fundamental ($\omega$, red) pulse is incident normally on the sample. The third harmonic (3ω, green) is detected in the reflected direction by a CCD camera, a spectrometer, or a photodiode connected to a lock-in amplifier.

To quantitatively analyze the THG effect, an equation for the THG intensity ($I_{3\omega}$), can be introduced [112]:

$$I_{3\omega}(t) = \frac{9\omega^2}{16|\tilde{n}_{3\omega}||\tilde{n}_\omega|^3 \epsilon_0^2 c^4} I_\omega^3 |\chi^{(3)}|^2 \left( \frac{e^{-2\alpha t} - 2\cos(\Delta k t)e^{-\alpha t} + 1}{\alpha^2 + \Delta k^2} \right) e^{-2\alpha t} \quad (6)$$

where $\tilde{n}_\omega$ and $\tilde{n}_{3\omega}$ are the complex refractive indexes at the fundamental and harmonic wavelengths, respectively, $\alpha$ is the absorption coefficient at the THG wavelength, $\Delta k$ is the phase mismatch between the fundamental and harmonic waves, and $\chi^{(3)}$ is the third-order susceptibility of the sample. By fitting the THG data with Equation (6), an effective third order susceptibility $\chi^{(3)}$ value can be obtained.

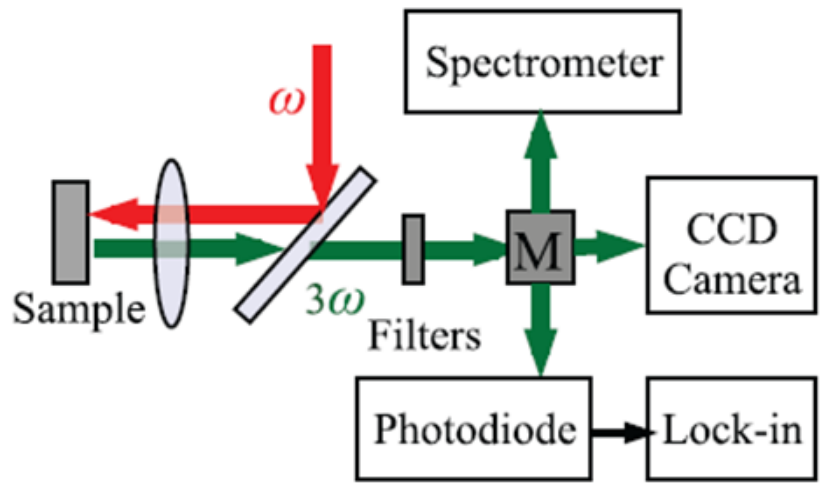

**Figure 5**. The scheme of a typical THG setup: CCD, charge-coupled-device [111].

Strong THG in graphene was demonstrated by Kumar et al.[111] Figure 6(a-i) shows the THG of monolayer graphene as a function of incident laser powers. The incident laser was 1720.4 nm. By fitting the experimental data, a large $\chi^{(3)}$ of ~ 0.4 × $10^{-16}$ m$^2$/V$^2$ was obtained. In addition, a thickness dependent THG signal can be observed (Figure 6(a-ii), while $\chi^{(3)}$ remains constant wth increasing graphene layer number. Recently, Jiang et al. [113] investigated the gate-tuneable THG of graphene. Figure 6(b-ii) shows the THG signal as a function of chemical potential generated at different wavelengths. When tuning the doping level of graphene, an enhanced THG and $\chi^{(3)}$ were observed.

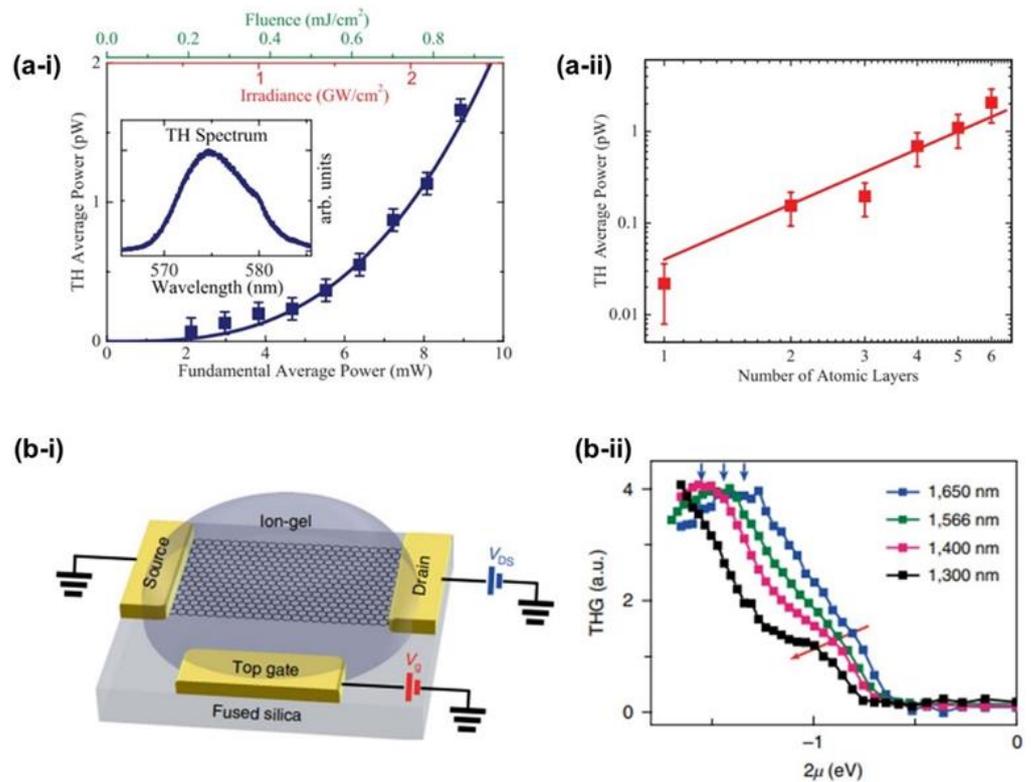

**Figure 6**. (a) THG in graphene:[111] (i) The average power of the THG signals as a function of the average power of the incident laser. Inset is the THG spectrum. (ii) The average power of the THG signals as a function of the number of atomic layers for an average fundamental power of 1 mW. (b) Gate-tunable THG in graphene:[113] (i) Schematic of an ion-gel-gated graphene monolayer on a fused silica substrate covered by ion-gel and voltage biased by the top gate. (ii) THG signal as a function of $2\mu$ generated by different input wavelengths: 1,300 nm, 1,400 nm, 1,566 nm and 1,650 nm.

THG in other 2D materials, such as TMDCs and BP, have also been investigated recently. Rosa et al.[114] characterized THG in mechanically exfoliated WSe$_2$ flakes at an excitation wavelength of 1560 nm. By measuring the THG for different numbers of layers, a clear thickness-dependent behaviour was observed, as shown in Figures 7(a-i) and (a-iii). The $\chi^{(3)}$ of WSe$_2$ was measured to be in the order of $10^{-19}$ m$^2$/V$^2$, which is comparable to other TMD [115] and BP [116]. Youngblood et al.[116] reported THG in BP by using an ultrafast near-IR laser obtaining a $\chi^{(3)}$ of $\sim 1.4 \times 10^{-19}$ m$^2$/V$^2$. In addition, an anisotropic THG were demonstrated, as shown in Figure 7(b-iii). Nonlinear optical properties of few-layer GaTe were also studied by characterizing the THG at a pump wavelength of 1560 nm [117]. The THG intensity was found to be sensitive to the number of GaTe layers (Figure 7(c-iii)). They obtained a large $\chi^{(3)}$ of $\sim 2 \times 10^{-16}$ m$^2$/V$^2$.

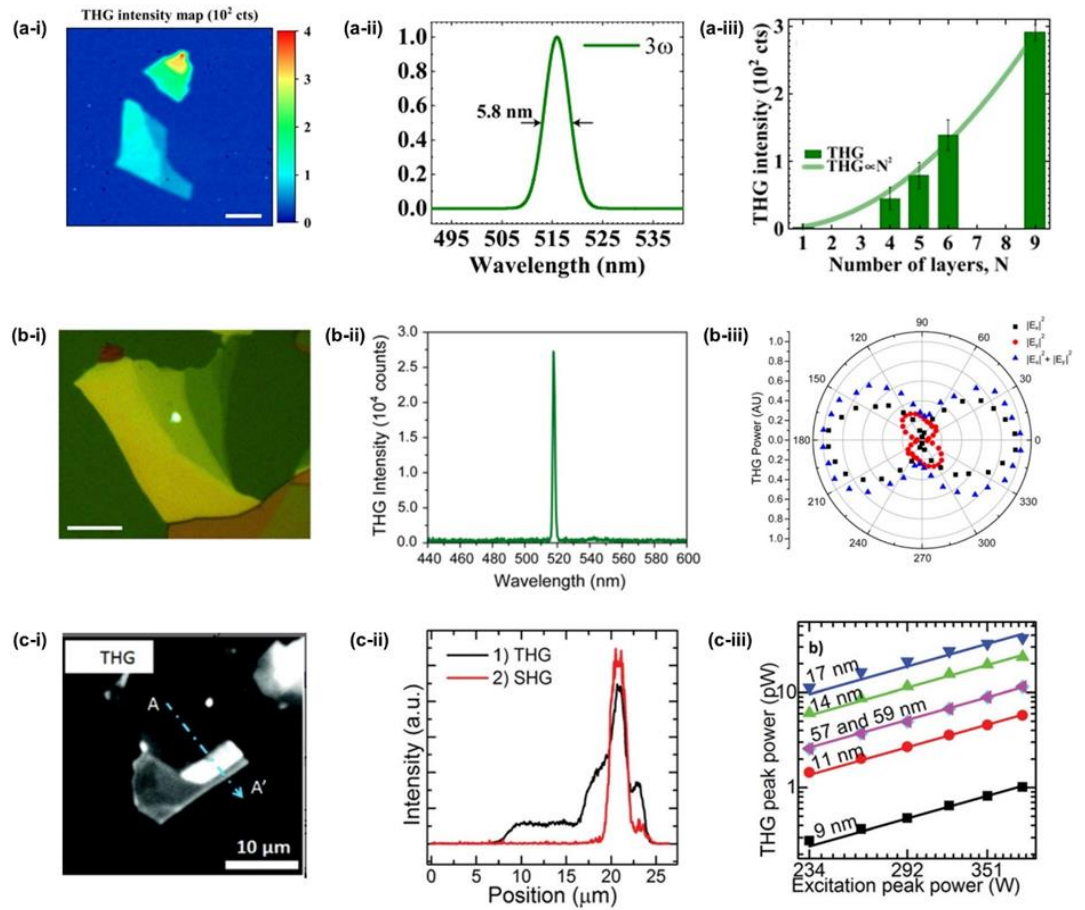

**Figure 7**. (a) THG in WSe$_2$:[114] (i) Spatial THG intensity mapping across the WSe$_2$ sample. (ii) THG spectrum of WSe$_2$. (iii) THG intensities as a function of sample layers. (b) THG in BP:[116] (i) THG emission (bright spot) from the BP flake. (ii) Measured spectrum of THG emission with a peak wavelength at 519 nm. (iii) Anisotropic THG in BP. (c) THG in GaTe:[117] (i) THG images of the few-layer GaTe flake. (ii) Measured spectra of THG emission. (iii) THG signals of samples with different thicknesses.

*3.2.3 Hybrid device characterization*

Z-scan and THG measurements are usually employed to characterize the material property directly. While on the one hand, the properties of a material form the basis for applications to electronic and optical devices, the reverse is true – device performance can also provide key information about the material properties. A typical example is field effect transistors (FETs) which have been one of the main techniques to evaluate the electrical properties of 2D materials. Optical structures and waveguides can also be exploited to characterize the material optical properties. By integrating 2D materials with photonic cavities and optical waveguides, the third-order optical nonlinearity of atomically thin 2D material has been characterized by measuring the nonlinear optical responses of the hybrid devices, such as FWM [78], SPM [99], and supercontinuum generation [118]. This method also enables the investigation of the layer-dependence of the nonlinear properties, which is challenging for conventional Z-scan methods due to the weak response of ultrathin 2D films.

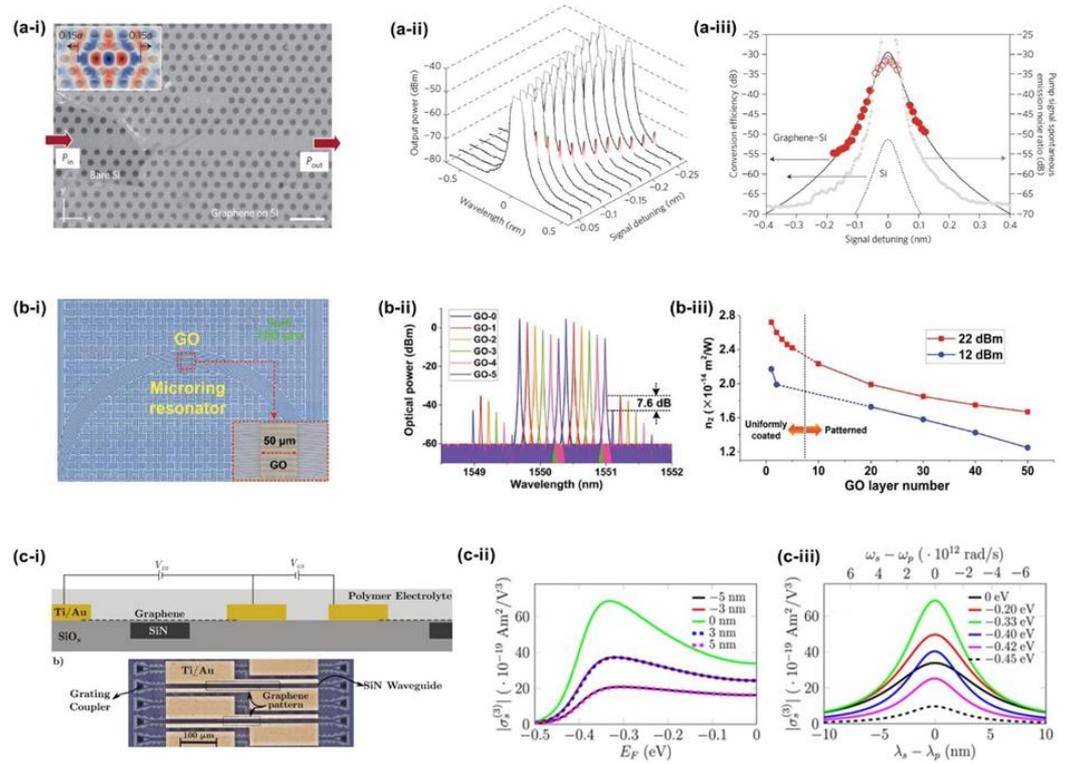

**Figure 8**. (a) FWM in graphene-clad silicon nanocavities:[118] (i) Scanning electron micrograph (SEM) of the photonic crystal cavity partially covered by graphene monolayer. (ii) Measured transmission spectrum of the cavity device with pump laser fixed on cavity resonance, and signal laser detuning scanned from 20.04 to 20.27 nm. (iii) Measured and simulated conversion efficiencies of the cavity. Solid and dashed black lines are modelled conversion efficiencies for the graphene–silicon and monolithic silicon cavities, respectively. (b) Layer dependent optical nonlinearities in GO-coated integrated MRR:[78] (i) Microscopic image of an integrated MRR patterned with 50 layers of GO. Inset shows zoom-in view of the patterned GO film. (ii) Optical spectra of FWM at a pump power of 22 dBm for the MRRs with 1–5 layers coated GO. (iii) $n_2$ of GO versus layer number at fixed pump powers of 12 and 22 dBm. (c) Electrically tunable optical nonlinearities in graphene-covered SiN Waveguides:[119] (i) Sketch of the gating scheme (up) and optical microscope image (down) of the device. Calculated values of third-order conductivity as a function of Fermi energy for different wavelength detunings (ii) and as a function of detuning for a range of Fermi energies (iii).

For the hybrid device characterization, the data analysis is performed in the following steps. First, by fitting the measured FWM or SPM spectra of corresponding hybrid devices, one can obtain the nonlinear parameters ($\gamma$) for the bare and hybrid waveguides. Then based on the fit $\gamma$ of the hybrid waveguides, the Kerr coefficient ($n_2$) of the coated 2D films can be extracted using [120-122]:

$$\gamma = \frac{2\pi}{\lambda} \frac{\iint_D n_0^2(x,y) n_2(x,y) S_z^2 dx dy}{\left[\iint_D n_0(x,y) S_z dx dy\right]^2} \tag{7}$$

where $\lambda$ is the central wavelength, $D$ is the integral of the optical fields over the material regions, $S_z$ is the time-averaged Poynting vector calculated using mode solving software, $n_0$ (*x*, *y*) and $n_2$ (*x*, *y*) are the refractive index profiles calculated over the waveguide cross section and the Kerr coefficient of the different material regions, respectively.

FWM is a fundamental third-order nonlinear optical process that has been widely used for all optical signal generation and processing, including wavelength conversion [98, 123], optical frequency comb generation [124, 125], optical sampling [126, 127], quantum entanglement [29, 30], and many other processes. The conversion efficiency (CE) of FWM is mainly determined by the third-order Kerr nonlinearity of the material that makes of the device. Therefore, it is useful to obtain the Kerr coefficient of a material by measuring its FWM CE.

Gu et al.[118] fabricated a silicon nanocavity covered with graphene (Figure 8(a-i)) and measured the FWM CE with different pump and signal detuning wavelengths around 1550 nm, as shown in Figures 8(a-ii) and (a-iii). From the CE data, a $n_2$ of ~ 4.8 × 10$^{-17}$ m$^2$/W was obtained for a graphene integrated with a silicon cavity. The layer-dependence of the Kerr nonlinearity of GO films has been investigated by measuring the FWM performance of GO hybrid devices based on doped-silica and SiN optical waveguides and microring resonators (MRRs) [78, 128-130]. Figure 8(b-i) shows a fabricated doped-silica MRR covered with patterned GO films.[78] By fitting the CE to theory for a device with different GO thicknesses, the layer thickness dependence of $n_2$ of GO at 1550 nm was characterized, as shown in Figure 8(b-iii). Recently, electrically tuneable optical nonlinearities of graphene at 1550 nm was also demonstrated by measuring FWM in graphene-SiN waveguides at different gate voltages, as shown in Figure 8(c) [119].

SPM is another third-order nonlinear optical process that can be used to characterize the optical nonlinearity of 2D materials. Feng et al.[131] studied the Kerr nonlinearities of graphene/Si hybrid waveguides with enhanced SPM (Figure 9(a)). The $n_2$ of the Graphene on Si hybrid waveguides was measured to be ~ 2 × 10$^{-17}$ m$^2$/W, which is three times larger than that of the Si waveguide. Even though the intrinsic $n_2$ of graphene is orders of magnitude larger than bulk silicon, the monolayer thickness of the graphene film results in a very low optical mode overlap, which yields only a factor of three improvement in the effective nonlinearity of the waveguide. For GO, on the other hand, comparatively larger film thicknesses are achievable which result in an overall much higher waveguide nonlinearity. Optical nonlinearities of GO films have also been investigated by SPM experiments. Zhang et al.[99] demonstrated the enhanced optical nonlinearity of silicon nanowires integrated with 2D GO Films (Figure 9(b-i)). Figure 9(b-ii) shows the experimental SPM spectra of the devices with different numbers of GO layers, where increased spectral broadening can be observed in GO coated silicon nanowires. The layer dependent Kerr $n_2$ coefficient of GO was also characterized by fitting the spectra to theory, as shown in Figure 9(b-iii). In addition to graphene and GO, the optical Kerr nonlinearity of MoS$_2$ monolayer films was also characterized by analysing the SPM of MoS$_2$-silicon waveguides [132]. The experiments demonstrated a large Kerr coefficient $n_2$ of ~ 1.1 × 10$^{-16}$ m$^2$/W for a monolayer of MoS$_2$ in the telecommunications band.

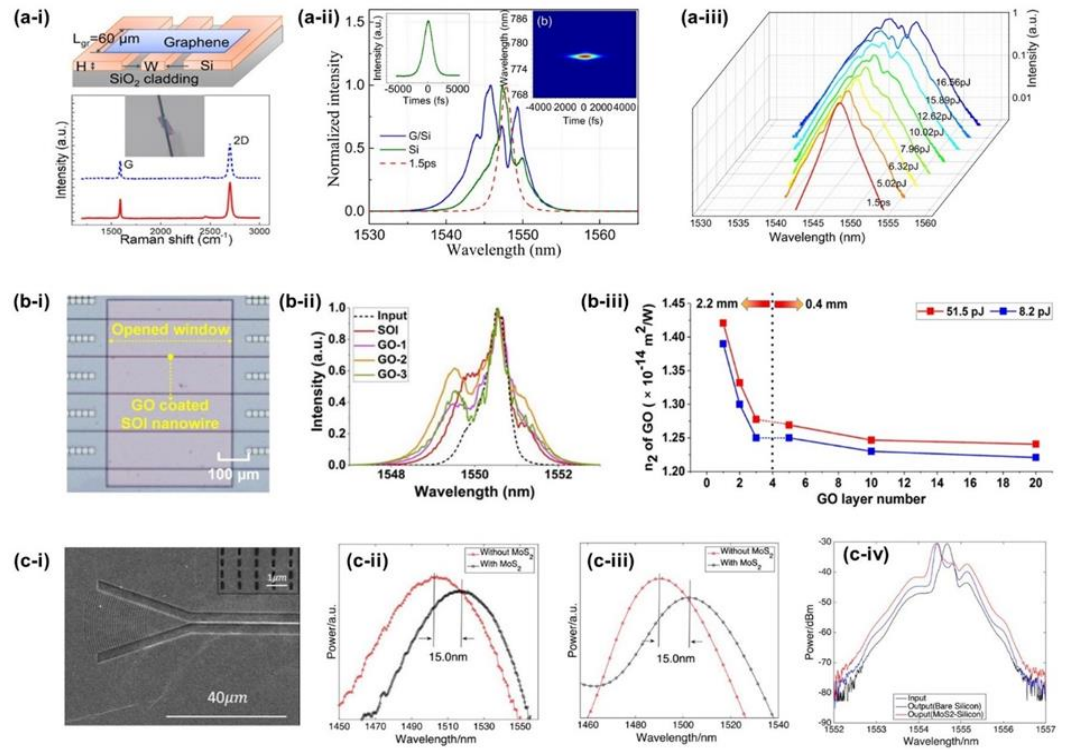

**Figure 9**. (a) SPM experiments in graphene-silicon hybrid waveguides:[131] (i) Schematic diagram and Raman spectra of the device. (ii) Measured transmission spectra of comparison between the Si (green solid curve) and G/Si hybrid (blue solid curve) waveguides under the same input energy with 1.5 ps input pulse (spectrum denoted by the red dashed curve). (iii) Output SPM spectra of the hybrid waveguide under various coupled energies. (b) SPM experiments in GO-silicon waveguide:[99] (i) Microscopic image of a GO-coated silicon nanowire. (ii) Optical spectra of SPM at a coupled pulse energy of~51.5 pJ with 1−3 layers coated GO. (iii) $n_2$ of GO vs layer number at fixed coupled pulse energies of 8.2 and 51.5 pJ. (c) SPM experiments in MoS$_2$-coated silicon waveguides:[132] (i) scanning electron microscope image of the device, with MoS$_2$ covering both the grating couplers and waveguide regions. (ii) Measured transmission spectra of the devices with and without MoS$_2$. (iii) Simulation result for the redshift of the grating to estimate the refractive index of MoS$_2$. (iv) SPM spectra of MoS$_2$–silicon waveguide and bare silicon waveguide.

### 3.3 Comparison of measured results

By using different characterization techniques discussed above, Kerr coefficient $n_2$ or THG $\chi^{(3)}$ value of graphene, GO, TMDCs, BP, and different heterostructures at telecommunication wavelengths have been obtained. In Table 1, we compare these parameters characterizing the third-order optical nonlinearity. It can be seen that monolayer graphene exhibits the largest $n_2$ value (up to $10^{-11}$ m$^2$/W). The $n_2$ value of GO films is on the magnitude of $10^{14}$ m$^2$/W, which is relatively smaller than graphene, but still more than 3 orders of magnitudes larger than that of bulk silicon. For MoS$_2$, MXene film, and 2D heterostructures, the measured $n_2$ varies from $10^{-16}$ to $10^{-22}$ m$^2$/W. In terms of $\chi^{(3)}$ susceptivity obtained by using THG measurements, the value ranges from $10^{-19}$ to $10^{-15}$ m$^2$/V$^2$ for graphene, TMDCs, and BP.

**Table 1.** Comparison of third-order optical nonlinear parameters of different 2D materials. FWM: four-wave mixing; SPM: self-phase modulation; WG: waveguide; MRR: microring resonator; THG: third-harmonic generation.

| Material | Wavelength [a] | Thickness | Nonlinear parameter | Method | Ref. |
|---|---|---|---|---|---|
| Graphene | 1550 nm | ~ 1 layer | $n_2 = \sim 10^{-11}$ m$^2$/W | Z-scan | [105] |
| Graphene | 1550 nm | ~ 5-7 layers | $n_2 = \sim -8 \times 10^{-14}$ m$^2$/W | Z-scan | [58] |
| GO | 1560 nm | ~ 1 um | $n_2 = \sim 4.5 \times 10^{-14}$ m$^2$/W | Z-scan | [60] |
| GeP | 1550 nm | ~ 15-40 nm | $n_2 = \sim 3.3 \times 10^{-19}$ m$^2$/W | Z-scan | [133] |
| CH$_3$NH$_3$PbI$_3$ | 1560 nm | ~ 180 nm | $n_2 = \sim 1.6 \times 10^{-12}$ m$^2$/W | Z-scan | [106] |
| MXene | 1550 nm | ~ 220 um | $n_2 = \sim -4.89 \times 10^{-20}$ m$^2$/W | Z-scan | [57] |
| MOF | 1550 nm | ~ 4.2 nm | $n_2 = \sim -8.9 \times 10^{-20}$ m$^2$/W | Z-scan | [134] |
| MoS$_2$/BP/MoS$_2$ | 1550 nm | ~ 17-20 nm | $n_2 = \sim 3.04 \times 10^{-22}$ m$^2$/W | Z-scan | [107] |
| Graphene/Bi$_2$Te$_3$ | 1550 nm | ~ 8.5 nm | $n_2 = \sim 2 \times 10^{-12}$ m$^2$/W | Z-scan | [110] |
| Graphene | 1550 nm | ~ 1 layer | $n_2 = \sim 10^{-13}$ m$^2$/W | SPM in WG | [135] |
| GO | 1550 nm | ~ 4 nm | $n_2 = \sim 1.5 \times 10^{-14}$ m$^2$/W | FWM in WG | [128] |
| GO | 1550 nm | ~ 2-100 nm | $n_2 = \sim (1.2-2.7) \times 10^{-14}$ m$^2$/W | FWM in MRR | [[78]] |
| GO | 1550 nm | ~ 2-20 nm | $n_2 = \sim (1.3-1.4) \times 10^{-14}$ m$^2$/W | FWM in WG | [129] |
| GO | 1550 nm | ~ 2-40 nm | $n_2 = \sim (1.2-1.4) \times 10^{-14}$ m$^2$/W | SPM in WG | [99] |
| MoS$_2$ | 1550 nm | ~ 1 layer | $n_2 = \sim 1.1 \times 10^{-16}$ m$^2$/W | SPM in WG | [132] |
| Graphene | 1560 nm | ~ 1 layer | $\chi^{(3)} = \sim 4 \times 10^{-15}$ m$^2$/V$^2$ | THG | [136] |
| Graphene | 1560 nm | ~ 1 layer | $\chi^{(3)} = \sim 1.5 \times 10^{-19}$ m$^2$/V$^2$ | THG | [137] |
| MoS$_2$ | 1560 nm | ~ 1 layer | $\chi^{(3)} = \sim 2.4 \times 10^{-19}$ m$^2$/V$^2$ | THG | [137] |
| MoSe$_2$ | 1560 nm | ~ 1 layer | $\chi^{(3)} = \sim 2.2 \times 10^{-19}$ m$^2$/V$^2$ | THG | [138] |
| WS$_2$ | 1560 nm | ~ 1 layer | $\chi^{(3)} = \sim 2.4 \times 10^{-19}$ m$^2$/V$^2$ | THG | [138] |
| WSe$_2$ | 1560 nm | ~ 1 layer | $\chi^{(3)} = \sim 1.2 \times 10^{-19}$ m$^2$/V$^2$ | THG | [114] |
| SnSe$_2$ | 1560 nm | multilayer | $\chi^{(3)} = \sim 4.1 \times 10^{-19}$ m$^2$/V$^2$ | THG | [139] |
| ReS$_2$ | 1515 nm | ~ 1 layer | $\chi^{(3)} = \sim 5.3 \times 10^{-18}$ m$^2$/V$^2$ | THG | [140] |
| BP | 1560 nm | multilayer | $\chi^{(3)} = \sim 1.6 \times 10^{-19}$ m$^2$/V$^2$ | THG | [141] |

[a] Here is the excitation laser wavelength.

## 4. Outlook and prospects

Despite these remarkable achievements, challenges still exist for engineering the nonlinear optical properties of 2D materials. First, accurate and efficient characterization of the linear and nonlinear optical properties remains challenging. Although the Z-scan method has been highly successful, the very weak Z-scan signals of ultra-thin films limit its applications in mono- or few layer 2D materials, especially for the layer-dependent measurements. In contrast, integrating 2D films with optical waveguides provides a powerful method to obtain accurate nonlinear parameters of atomic-thin 2D materials by analyzing the nonlinear optical performance of the hybrid device. However, the complicated device fabrication process and resulting relatively low efficiency make this method unsuitable for rapid material characterization which is required for future industrial applications. Second, 2D materials are a large family which include thousands of different materials. For applications of the third-order optical nonlinearity in the telecommunications band, only a very small fraction of them have been investigated. Many newer materials, such as perovskites, MOFs, and graphdiyne, still need more research, which hinders the full exploitation of 2D materials in the fabrication of next-generation nonlinear optical devices. Finally, tuning or engineering the properties of materials is important for both optimizing the device performance and enabling new functionalities, as well as the fundamental study of 2D materials. Nevertheless, current advances in the study of third-order optical nonlinearities of 2D materials focus mainly on their fundamental properties. The relative lack of effective methods of tuning the material properties poses another obstacle for 2D materials to move forward to practical device fabrication. While challenges remain and more work is needed, there is no doubt that 2D materials will underpin key breakthroughs and greatly accelerate the developments of next-generation nonlinear optical devices for many applications, particularly high bandwidth optical communications systems.

## 5. Conclusions

In conclusion, we review recent progress in the study of the third-order optical nonlinearities of 2D materials in the telecommunications wavelength band. We introduce the representative 2D materials, together with their basic material properties followed by a discussion of the main methods for characterizing the third-order optical nonlinearity, reviewing recent achievements in the field. These advances highlight the significant potential of 2D materials in enabling high-performance nonlinear optical devices for all-optical processing functions in optical communications systems.

**Funding:** Please add: This research received no external funding.

**Data Availability Statement:** Not applicable.

**Conflicts of Interest:** The authors declare no conflict of interest.

## References

1. Leuthold, J., C. Koos, and W. Freude, *Nonlinear silicon photonics.* Nature Photonics, 2010. **4**(8): p. 535-544.
2. Xia, F., L. Sekaric, and Y. Vlasov, *Ultracompact optical buffers on a silicon chip.* Nature Photonics, 2007. **1**(1): p. 65-71.
3. Foster, M.A., et al., *Broad-band optical parametric gain on a silicon photonic chip.* Nature, 2006. **441**(7096): p. 960-3.
4. Razzari, L., et al., *CMOS-compatible integrated optical hyper-parametric oscillator.* Nature Photonics, 2009. **4**(1): p. 41-45.
5. Ferrera, M., et al., *Low-power continuous-wave nonlinear optics in doped silica glass integrated waveguide structures.* Nature Photonics, 2008. **2**(12): p. 737-740.
6. Gaeta, A.L., M. Lipson, and T.J. Kippenberg, *Photonic-chip-based frequency combs.* Nature Photonics, 2019. **13**(3): p. 158-169.
7. Demongodin, P., et al., *Ultrafast saturable absorption dynamics in hybrid graphene/Si3N4 waveguides.* APL Photonics, 2019. **4**(7): p. 076102.
8. Aitchison, J.S., et al., *The nonlinear optical properties of AlGaAs at the half band gap.* IEEE Journal of Quantum Electronics, 1997. **33**(3): p. 341-348.
9. Kawashima, H., et al., *Optical Bistable Response in AlGaAs-Based Photonic Crystal Microcavities and Related Nonlinearities.* IEEE Journal of Quantum Electronics, 2008. **44**(9): p. 841-849.
10. Xie, W., et al., *Ultrahigh-Q AlGaAs-on-insulator microresonators for integrated nonlinear photonics.* Opt Express, 2020. **28**(22): p. 32894-32906.
11. Nicoletti, E., et al., *Generation of lambda/12 nanowires in chalcogenide glasses.* Nano Lett, 2011. **11**(10): p. 4218-21.
12. Eggleton, B.J., B. Luther-Davies, and K. Richardson, *Chalcogenide photonics.* Nature Photonics, 2011. **5**(3): p. 141-148.
13. Liang, T.K. and H.K. Tsang, *Efficient Raman amplification in silicon-on-insulator waveguides.* Applied Physics Letters, 2004. **85**(16): p. 3343-3345.
14. Yamashita, D., et al., *Strongly asymmetric wavelength dependence of optical gain in nanocavity-based Raman silicon lasers.* Optica, 2018. **5**(10): p. 1256.
15. Zhang, Y., K. Zhong, and H.K. Tsang, *Raman Lasing in Multimode Silicon Racetrack Resonators.* Laser & Photonics Reviews, 2020. **15**(2): p. 2000336.
16. Salem, R., et al., *Signal regeneration using low-power four-wave mixing on silicon chip.* Nature Photonics, 2008. **2**(1): p. 35-38.
17. Mathlouthi, W., H. Rong, and M. Paniccia, *Characterization of efficient wavelength conversion by four-wave mixing in sub-micron silicon waveguides.* Optics Express, 2008. **16**(21): p. 16735-16745.
18. Frigg, A., et al., *Optical frequency comb generation with low temperature reactive sputtered silicon nitride waveguides.* APL Photonics, 2020. **5**(1): p. 011302.
19. Li, F., et al., *All-optical XOR logic gate for 40Gb/s DPSK signals via FWM in a silicon nanowire.* Optics Express, 2011. **19**(21): p. 20364-20371.
20. Gao, S., et al., *Reconfigurable dual-channel all-optical logic gate in a silicon waveguide using polarization encoding.* Optics Letters, 2015. **40**(7): p. 1448-1451.
21. Soref, R., et al., *Silicon-Based Group-IV O-E-O Devices for Gain, Logic, and Wavelength Conversion.* ACS Photonics, 2020. **7**(3): p. 800-811.
22. Monat, C., et al., *Integrated optical auto-correlator based on third-harmonic generation in a silicon photonic crystal waveguide.* Nat Commun, 2014. **5**: p. 3246.
23. Wu, J., et al., *RF Photonics: An Optical Microcombs' Perspective.* IEEE Journal of Selected Topics in Quantum Electronics, 2018. **24**(4): p. 1-20.
24. Xu, X., et al., *Broadband RF Channelizer Based on an Integrated Optical Frequency Kerr Comb Source.* Journal of Lightwave Technology, 2018. **36**(19): p. 4519-4526.
25. Xu, X., et al., *Microcomb-Based Photonic RF Signal Processing.* IEEE Photonics Technology Letters, 2019. **31**(23): p. 1854-1857.

26. Corcoran, B., et al., *Ultra-dense optical data transmission over standard fibre with a single chip source.* Nat Commun, 2020. **11**(1): p. 2568.
27. Fridman, M., et al., *Demonstration of temporal cloaking.* Nature, 2012. **481**(7379): p. 62-65.
28. Christian Reimer, M.K., Piotr Roztocki, Benjamin Wetzel, Fabio Grazioso, Brent E. Little, Sai T. Chu, Tudor Johnston, Yaron Bromberg, Lucia Caspani, David J. Moss, Roberto Morandotti,, *Generation of multiphoton entangled quantum states by means of integrated frequency combs.* Science, 2016. **351**(6278): p. 4.
29. Kues, M., et al., *On-chip generation of high-dimensional entangled quantum states and their coherent control.* Nature, 2017. **546**(7660): p. 622-626.
30. Caspani, L., et al., *Integrated sources of photon quantum states based on nonlinear optics.* Light Sci Appl, 2017. **6**(11): p. e17100.
31. Geim, A.K. and K.S. Novoselov, *The rise of graphene.* Nature materials, 2007. **6**(3): p. 183-191.
32. Yang, T., et al., *Tailoring pores in graphene-based materials: from generation to applications.* Journal of Materials Chemistry A, 2017. **5**(32): p. 16537-16558.
33. Bolotin, K.I., et al., *Temperature-dependent transport in suspended graphene.* Phys Rev Lett, 2008. **101**(9): p. 096802.
34. Loh, K.P., et al., *Graphene oxide as a chemically tunable platform for optical applications.* Nat Chem, 2010. **2**(12): p. 1015-24.
35. Ghofraniha, N. and C. Conti, *Graphene oxide photonics.* Journal of Optics, 2019. **21**(5): p. 053001.
36. Luo, Z., et al., *Photoluminescence and band gap modulation in graphene oxide.* Applied Physics Letters, 2009. **94**(11): p. 111909.
37. Tian, H., et al., *Optoelectronic devices based on two-dimensional transition metal dichalcogenides.* Nano Research, 2016. **9**(6): p. 1543-1560.
38. Tan, T., et al., *2D Material Optoelectronics for Information Functional Device Applications: Status and Challenges.* Adv Sci, 2020. **7**(11): p. 2000058.
39. Liang, Q., et al., *High-Performance, Room Temperature, Ultra-Broadband Photodetectors Based on Air-Stable $PdSe_2$.* Adv Mater, 2019. **31**(24): p. e1807609.
40. Pi, L., et al., *Recent Progress on 2D Noble‐Transition‐Metal Dichalcogenides.* Advanced Functional Materials, 2019. **29**(51): p. 1904932.
41. Zhang, K., et al., *Two dimensional hexagonal boron nitride (2D-hBN): synthesis, properties and applications.* Journal of Materials Chemistry C, 2017. **5**(46): p. 11992-12022.
42. Tran, T.T., et al., *Quantum emission from hexagonal boron nitride monolayers.* Nat Nanotechnol, 2016. **11**(1): p. 37-41.
43. Caldwell, J.D., et al., *Photonics with hexagonal boron nitride.* Nature Reviews Materials, 2019. **4**(8): p. 552-567.
44. Autere, A., et al., *Nonlinear Optics with 2D Layered Materials.* Adv Mater, 2018. **30**(24): p. 1705963.
45. Qiao, J., et al., *High-mobility transport anisotropy and linear dichroism in few-layer black phosphorus.* Nature Communications, 2014. **5**(1): p. 4475.
46. Yuan, H., et al., *Polarization-sensitive broadband photodetector using a black phosphorus vertical p–n junction.* Nature Nanotechnology, 2015. **10**(8): p. 707-713.
47. Guo, Q., et al., *Black Phosphorus Mid-Infrared Photodetectors with High Gain.* Nano Letters, 2016. **16**(7): p. 4648-4655.
48. Li, L., et al., *Direct observation of the layer-dependent electronic structure in phosphorene.* Nat Nanotechnol, 2017. **12**(1): p. 21-25.
49. Radisavljevic, B., et al., *Single-layer $MoS_2$ transistors.* Nat Nanotechnol, 2011. **6**(3): p. 147-50.
50. Fang, H., et al., *High performance single layered $WSe_2$ p-FETs with chemically doped contacts.* Nano Lett, 2012. **12**(7): p. 3788-92.
51. Michael M. Lee, J.T., Tsutomu Miyasaka, Takurou N. Murakami, and Henry J. Snaith, *Efficient Hybrid Solar Cells Based on Meso-Superstructured Organometal Halide Perovskites.* Science, 2012. **338**(6107): p. 643-647.
52. Xing, J., et al., *High-Efficiency Light-Emitting Diodes of Organometal Halide Perovskite Amorphous Nanoparticles.* ACS Nano, 2016. **10**(7): p. 6623-30.

53. Zhang, J., et al., *Thickness-dependent nonlinear optical properties of CsPbBr3 perovskite nanosheets.* Opt Lett, 2017. **42**(17): p. 3371-3374.
54. Sun, Z., et al., *Graphene Mode-Locked Ultrafast Laser.* ACS nano, 2010. **4**(2): p. 803-810.
55. Zhang, J., et al., *Ultrafast saturable absorption of MoS2 nanosheets under different pulse-width excitation conditions.* Opt Lett, 2018. **43**(2): p. 243-246.
56. Liu, Y., et al., *Investigation of mode coupling in normal-dispersion silicon nitride microresonators for Kerr frequency comb generation.* Optica, 2014. **1**(3): p. 137.
57. Jiang, X., et al., *Broadband Nonlinear Photonics in Few-Layer MXene Ti3C2Tx (T = F, O, or OH).* Laser & Photonics Reviews, 2018. **12**(2): p. 1700229.
58. Demetriou, G., et al., *Nonlinear optical properties of multilayer graphene in the infrared.* Opt Express, 2016. **24**(12): p. 13033-43.
59. Zheng, X., et al., *In situ third-order non-linear responses during laser reduction of graphene oxide thin films towards on-chip nonlinear photonic devices.* Adv Mater, 2014. **26**(17): p. 2699-703.
60. Xu, X., et al., *Observation of Third-order Nonlinearities in Graphene Oxide Film at Telecommunication Wavelengths.* Sci Rep, 2017. **7**(1): p. 9646.
61. Jia, L., et al., *Highly nonlinear BiOBr nanoflakes for hybrid integrated photonics.* APL Photonics, 2019. **4**(9): p. 090802.
62. 62. Jia, L.N., et al., *Large Third-Order Optical Kerr Nonlinearity in Nanometer-Thick PdSe2 2D Dichalcogenide Films: Implications for Nonlinear Photonic Devices.* Acs Applied Nano Materials, 2020. **3**(7): p. 6876-6883.
63. Yoshikawa, N., T. Tamaya, and K. Tanaka, *High-harmonic generation in graphene enhanced by elliptically polarized light excitation.* Science, 2017. **356**(6339): p. 736-738.
64. Janisch, C., et al., *Extraordinary Second Harmonic Generation in tungsten disulfide monolayers.* Sci Rep, 2014. **4**: p. 5530.
65. Youngblood, N., et al., *Layer-Tunable Third-Harmonic Generation in Multilayer Black Phosphorus.* ACS Photonics, 2016. **4**(1): p. 8-14.
66. Jiang, T., et al., *Ultrafast coherent nonlinear nanooptics and nanoimaging of graphene.* Nat Nanotechnol, 2019. **14**(9): p. 838-843.
67. Foster, M.A., et al., *Silicon-chip-based ultrafast optical oscilloscope.* Nature, 2008. **456**(7218): p. 81-4.
68. Zhong, H.-S., et al., *Quantum computational advantage using photons.* Science, 2020. **370**(6523): p. 1460.
69. Chen, W., et al., *Nonlinear Photonics Using Low-Dimensional Metal-Halide Perovskites: Recent Advances and Future Challenges.* Adv Mater, 2021. **33**(11): p. e2004446.
70. Liu, W., et al., *Recent Advances of 2D Materials in Nonlinear Photonics and Fiber Lasers.* Advanced Optical Materials, 2020. **8**(8): p. 1901631.
71. Nair, R.R., et al., *Fine Structure Constant Defines Visual Transparency of Graphene.* Science, 2008. **320**(5881): p. 1308.
72. Lui, C.H., et al., *Ultrafast Photoluminescence from Graphene.* Physical Review Letters, 2010. **105**(12): p. 127404.
73. Sun, Z., A. Martinez, and F. Wang, *Optical modulators with 2D layered materials.* Nature Photonics, 2016. **10**(4): p. 227-238.
74. Li, J.Z.L.L.F., *Graphene Oxide-Physics and Applications.* 2015.
75. Bao, Q., et al., *Atomic-Layer Graphene as a Saturable Absorber for Ultrafast Pulsed Lasers.* Advanced Functional Materials, 2009. **19**(19): p. 3077-3083.
76. Zhu, G., et al., *Graphene Mode-Locked Fiber Laser at 2.8 $\mu\text{m}$.* IEEE Photonics Technology Letters, 2016. **28**(1): p. 7-10.
77. Wu, R., et al., *Purely coherent nonlinear optical response in solution dispersions of graphene sheets.* Nano Lett, 2011. **11**(12): p. 5159-64.
78. Wu, J., et al., *2D Layered Graphene Oxide Films Integrated with Micro-Ring Resonators for Enhanced Nonlinear Optics.* Small, 2020. **16**(16): p. e1906563.
79. Hendry, E., et al., *Coherent nonlinear optical response of graphene.* Phys Rev Lett, 2010. **105**(9): p. 097401.

80. Splendiani, A., et al., *Emerging photoluminescence in monolayer MoS2.* Nano Lett, 2010. **10**(4): p. 1271-5.
81. Mak, K.F., D. Xiao, and J. Shan, *Light–valley interactions in 2D semiconductors.* Nature Photonics, 2018. **12**(8): p. 451-460.
82. Xu, N., et al., *Palladium diselenide as a direct absorption saturable absorber for ultrafast mode-locked operations: from all anomalous dispersion to all normal dispersion.* Nanophotonics, 2020. **0**(0).
83. Yang, T., et al., *Anisotropic Third-Order Nonlinearity in Pristine and Lithium Hydride Intercalated Black Phosphorus.* ACS Photonics, 2018. **5**(12): p. 4969-4977.
84. Zhao, Y., et al., *Recent advance in black phosphorus: Properties and applications.* Materials Chemistry and Physics, 2017. **189**: p. 215-229.
85. Wu, H.-Y., Y. Yen, and C.-H. Liu, *Observation of polarization and thickness dependent third-harmonic generation in multilayer black phosphorus.* Applied Physics Letters, 2016. **109**(26).
86. Wang, Y., et al., *Ultrafast recovery time and broadband saturable absorption properties of black phosphorus suspension.* Applied Physics Letters, 2015. **107**(9): p. 091905.
87. Wang, K., et al., *Ultrafast Nonlinear Excitation Dynamics of Black Phosphorus Nanosheets from Visible to Mid-Infrared.* ACS Nano, 2016. **10**(7): p. 6923-32.
88. Luo, Z.C., et al., *Microfiber-based few-layer black phosphorus saturable absorber for ultra-fast fiber laser.* Opt Express, 2015. **23**(15): p. 20030-9.
89. Wang, J., F. Ma, and M. Sun, *Graphene, hexagonal boron nitride, and their heterostructures: properties and applications.* RSC Advances, 2017. **7**(27): p. 16801-16822.
90. Naguib, M., et al., *Two-dimensional nanocrystals produced by exfoliation of Ti3 AlC2.* Adv Mater, 2011. **23**(37): p. 4248-53.
91. Dillon, A.D., et al., *Highly Conductive Optical Quality Solution-Processed Films of 2D Titanium Carbide.* Advanced Functional Materials, 2016. **26**(23): p. 4162-4168.
92. Stranks, S.D. and H.J. Snaith, *Metal-halide perovskites for photovoltaic and light-emitting devices.* Nature Nanotechnology, 2015. **10**(5): p. 391-402.
93. Gu, C., et al., *Giant and Multistage Nonlinear Optical Response in Porphyrin-Based Surface-Supported Metal-Organic Framework Nanofilms.* Nano Lett, 2019. **19**(12): p. 9095-9101.
94. Liu, W., et al., *Structural Engineering of Low-Dimensional Metal-Organic Frameworks: Synthesis, Properties, and Applications.* Adv Sci, 2019. **6**(12): p. 1802373.
95. Medishetty, R., et al., *Nonlinear optical properties, upconversion and lasing in metal-organic frameworks.* Chem Soc Rev, 2017. **46**(16): p. 4976-5004.
96. Zheng, Y., et al., *Recent Progress in 2D Metal‐Organic Frameworks for Optical Applications.* Advanced Optical Materials, 2020. **8**(13): p. 2000110.
97. Boyd, R.W., *Nonlinear Optics.* Elsevier, Rochester, NY, USA, 2007.
98. Moss, D.J., et al., *New CMOS-compatible platforms based on silicon nitride and Hydex for nonlinear optics.* Nature Photonics, 2013. **7**(8): p. 597-607.
99. Zhang, Y., et al., *Enhanced Kerr Nonlinearity and Nonlinear Figure of Merit in Silicon Nanowires Integrated with 2D Graphene Oxide Films.* ACS Appl Mater Interfaces, 2020. **12**(29): p. 33094-33103.
100. Li, Y., et al., *All-optical RF spectrum analyzer with a 5 THz bandwidth based on CMOS-compatible high-index doped silica waveguides.* Optics Letters, 2021. **46**(7): p. 1574.
101. 101.        Ferrera, M., et al., *CMOS compatible integrated all-optical radio frequency spectrum analyzer.* Opt Express, 2014. **22**(18): p. 21488-98.
102. Corcoran, B., et al., *Green light emission in silicon through slow-light enhanced third-harmonic generation in photonic-crystal waveguides.* Nature Photonics, 2009. **3**(4): p. 206-210.

103. Corcoran, B., et al., *Optical signal processing on a silicon chip at 640Gb/s using slow-light.* Optics Express, 2010. **18**(8): p. 7770-7781.
104. Mansoor Sheik-Bahae, A.A.S., Tai-Huei Wei, David J. Hagan and E. W. Van Stryland, *Sensitive Measurement of Optical Nonlinearities Using a Single Beam.* IEEE JOURNAL OF QUANTUM ELECTRONICS, 1990. **26**(4): p. 10.
105. Zhang, H., et al., *Z-scan measurement of the nonlinear refractive index of graphene.* Optics Letters, 2012. **37**(11): p. 1856-1858.
106. Yi, J., et al., *Third-order nonlinear optical response of $CH_3NH_3PbI_3$ perovskite in the mid-infrared regime.* Optical Materials Express, 2017. **7**(11): p. 3894.
107. Xiao, S., et al., *Nonlinear optical modulation of MoS2/black phosphorus/MoS2 at 1550 nm.* Physica B: Condensed Matter, 2020. **594**: p. 412364.
108. Novoselov, K.S., et al., *2D materials and van der Waals heterostructures.* Science, 2016. **353**(6298): p. aac9439.
109. Liu, Y., et al., *Van der Waals heterostructures and devices.* Nature Reviews Materials, 2016. **1**(9): p. 16042.
110. Wang, Y., et al., *Observation of large nonlinear responses in a graphene-Bi2Te3 heterostructure at a telecommunication wavelength.* Applied Physics Letters, 2016. **108**(22): p. 221901.
111. Kumar, N., et al., *Third harmonic generation in graphene and few-layer graphite films.* Physical Review B, 2013. **87**(12): p. 121406.
112. Abdelwahab, I., et al., *Highly Enhanced Third-Harmonic Generation in 2D Perovskites at Excitonic Resonances.* ACS Nano, 2018. **12**(1): p. 644-650.
113. Jiang, T., et al., *Gate-tunable third-order nonlinear optical response of massless Dirac fermions in graphene.* Nature Photonics, 2018. **12**(7): p. 430-436.
114. Rosa, H.G., et al., *Characterization of the second- and third-harmonic optical susceptibilities of atomically thin tungsten diselenide.* Sci Rep, 2018. **8**(1): p. 10035.
115. Wang, R., et al., *Third-Harmonic Generation in Ultrathin Films of MoS2.* ACS Applied Materials & Interfaces, 2014. **6**(1): p. 314-318.
116. Youngblood, N., et al., *Layer-Tunable Third-Harmonic Generation in Multilayer Black Phosphorus.* ACS Photonics, 2017. **4**(1): p. 8-14.
117. Susoma, J., et al., *Second and third harmonic generation in few-layer gallium telluride characterized by multiphoton microscopy.* Applied Physics Letters, 2016. **108**(7): p. 073103.
118. Gu, T., et al., *Regenerative oscillation and four-wave mixing in graphene optoelectronics.* Nature Photonics, 2012. **6**(8): p. 554-559.
119. Alexander, K., et al., *Electrically Tunable Optical Nonlinearities in Graphene-Covered SiN Waveguides Characterized by Four-Wave Mixing.* ACS Photonics, 2017. **4**(12): p. 3039-3044.
120. Yang, Y., et al., *Invited Article: Enhanced four-wave mixing in waveguides integrated with graphene oxide.* APL Photonics, 2018. **3**(12): p. 120803.
121. Donnelly, C. and D.T. Tan, *Ultra-large nonlinear parameter in graphene-silicon waveguide structures.* Optics express, 2014. **22**(19): p. 22820-22830.
122. Ji, M., et al., *Enhanced parametric frequency conversion in a compact silicon-graphene microring resonator.* Optics Express, 2015. **23**(14): p. 18679-18685.
123. Pasquazi, A., et al., *All-optical wavelength conversion in an integrated ring resonator.* Optics Express, 2010. **18**(4): p. 3858-3863.

124. Xu, X., et al., *11 TOPS photonic convolutional accelerator for optical neural networks.* Nature, 2021. **589**(7840): p. 44-51.
125. Pasquazi, A., et al., *Micro-combs: A novel generation of optical sources.* Physics Reports, 2018. **729**: p. 1-81.
126. Koos, C., et al., *All-optical high-speed signal processing with silicon–organic hybrid slot waveguides.* Nature Photonics, 2009. **3**(4): p. 216-219.

127. Ji, H., et al., *Optical Waveform Sampling and Error-Free Demultiplexing of 1.28 Tb/s Serial Data in a Nanoengineered Silicon Waveguide.* Journal of Lightwave Technology, 2011. **29**(4): p. 426-431.
128. Qu, Y., et al., *Enhanced Four‐Wave Mixing in Silicon Nitride Waveguides Integrated with 2D Layered Graphene Oxide Films.* Advanced Optical Materials, 2020. **8**(23): p. 2001048.
129. Qu, Y., et al., *Analysis of Four-Wave Mixing in Silicon Nitride Waveguides Integrated With 2D Layered Graphene Oxide Films.* Journal of Lightwave Technology, 2021. **39**(9): p. 2902-2910.
130. Feng, Q., et al., *Enhanced optical Kerr nonlinearity of graphene/Si hybrid waveguide.* Applied Physics Letters, 2019. **114**(7).
131. Liu, L., et al., *Enhanced optical Kerr nonlinearity of MoS_2 on silicon waveguides.* Photonics Research, 2015. **3**(5): p. 206-209.
132. Guo, J., et al., *2D GeP as a Novel Broadband Nonlinear Optical Material for Ultrafast Photonics.* Laser & Photonics Reviews, 2019.
133. Jiang, X., et al., *Ultrathin Metal–Organic Framework: An Emerging Broadband Nonlinear Optical Material for Ultrafast Photonics.* Advanced Optical Materials, 2018. **6**(16): p. 1800561.
134. Vermeulen, N., et al., *Negative Kerr Nonlinearity of Graphene as seen via Chirped-Pulse-Pumped Self-Phase Modulation.* Physical Review Applied, 2016. **6**(4): p. 044006.
135. Ullah, K., et al., *Harmonic Generation in Low-Dimensional Materials.* Advanced Optical Materials, 2022. **10**(7): p. 2101860.
136. Woodward, R.I., et al., *Characterization of the second- and third-order nonlinear optical susceptibilities of monolayer MoS2 using multiphoton microscopy.* 2D Materials, 2017. **4**(1): p. 011006.
137. Autere, A., et al., *Optical harmonic generation in monolayer group-VI transition metal dichalcogenides.* Physical Review B, 2018. **98**(11): p. 115426.
138. Biswas, R., et al., *Third-harmonic generation in multilayer Tin Diselenide under the influence of Fabry-Perot interference effects.* Optics Express, 2019. **27**(20): p. 28855-28865.
139. Cui, Q., et al., *Strong and anisotropic third-harmonic generation in monolayer and multilayer ReS2.* Physical Review B, 2017. **95**(16): p. 165406.
140. Autere, A., et al., *Rapid and Large-Area Characterization of Exfoliated Black Phosphorus Using Third-Harmonic Generation Microscopy.* The Journal of Physical Chemistry Letters, 2017. **8**(7): p. 1343-1350.

141. Xu, X., et al., Photonic microwave true time delays for phased array antennas using a 49 GHz FSR integrated micro-comb source, *Photonics Research*, **6**, B30-B36 (2018).
142. M. Tan et al, "Orthogonally polarized Photonic Radio Frequency single sideband generation with integrated micro-ring resonators", IOP Journal of Semiconductors, Vol. **42** (4), 041305 (2021). DOI: 10.1088/1674-4926/42/4/041305.
143. Mengxi Tan, X. Xu, J. Wu, T. G. Nguyen, S. T. Chu, B. E. Little, R. Morandotti, A. Mitchell, and David J. Moss, "Photonic Radio Frequency Channelizers based on Kerr Optical Micro-combs", IOP Journal of Semiconductors Vol. **42** (4), 041302 (2021). DOI:10.1088/1674-4926/42/4/041302.
144. Xu, *et al.*, "Advanced adaptive photonic RF filters with 80 taps based on an integrated optical micro-comb source," *Journal of Lightwave Technology,* vol. 37, no. 4, pp. 1288-1295 (2019).
145. X. Xu, *et al.*, Broadband microwave frequency conversion based on an integrated optical micro-comb source", *Journal of Lightwave Technology*, vol. 38 no. 2, pp. 332-338, 2020.
146. M. Tan, *et al.*, "Photonic RF and microwave filters based on 49GHz and 200GHz Kerr microcombs", *Optics Comm.* vol. 465,125563, Feb. 22. 2020.
147. X. Xu, *et al.*, "Broadband photonic RF channelizer with 90 channels based on a soliton crystal microcomb", *Journal of Lightwave Technology*, Vol. 38, no. 18, pp. 5116 - 5121, 2020. doi: 10.1109/JLT.2020.2997699.

148. X. Xu, *et al.,* "Photonic RF and microwave integrator with soliton crystal microcombs", *IEEE Transactions on Circuits and Systems II: Express Briefs*, vol. 67, no. 12, pp. 3582-3586, 2020. DOI:10.1109/TCSII.2020.2995682.

149. X. Xu, *et al.*, "High performance RF filters via bandwidth scaling with Kerr micro-combs," *APL Photonics,* vol. 4 (2) 026102. 2019.

150. M. Tan, *et al.,* "Microwave and RF photonic fractional Hilbert transformer based on a 50 GHz Kerr micro-comb", *Journal of Lightwave Technology*, vol. 37, no. 24, pp. 6097 – 6104, 2019.

151. M. Tan, *et al.,* "RF and microwave fractional differentiator based on photonics", *IEEE Transactions on Circuits and Systems: Express Briefs*, vol. 67, no.11, pp. 2767-2771, 2020. DOI:10.1109/TCSII.2020.2965158.

152. M. Tan, *et al*., "Photonic RF arbitrary waveform generator based on a soliton crystal micro-comb source", Journal of Lightwave Technology, vol. 38, no. 22, pp. 6221-6226 (2020). DOI: 10.1109/JLT.2020.3009655.

153. M. Tan, X. Xu, J. Wu, R. Morandotti, A. Mitchell, and D. J. Moss, "RF and microwave high bandwidth signal processing based on Kerr Micro-combs", Advances in Physics X, VOL. 6, NO. 1, 1838946 (2021). DOI:10.1080/23746149.2020.1838946.

154. X. Xu, et al., "Advanced RF and microwave functions based on an integrated optical frequency comb source," Opt. Express, vol. 26 (3) 2569 (2018).

155. M. Tan, X. Xu, J. Wu, B. Corcoran, A. Boes, T. G. Nguyen, S. T. Chu, B. E. Little, R.Morandotti, A. Lowery, A. Mitchell, and D. J. Moss, ""Highly Versatile Broadband RF Photonic Fractional Hilbert Transformer Based on a Kerr Soliton Crystal Microcomb", Journal of Lightwave Technology vol. 39 (24) 7581-7587 (2021).

156. T. G. Nguyen *et al.*, "Integrated frequency comb source-based Hilbert transformer for wideband microwave photonic phase analysis," *Opt. Express,* vol. 23, no. 17, pp. 22087-22097, Aug. 2015.

157. X. Xu, J. Wu, M. Shoeiby, T. G. Nguyen, S. T. Chu, B. E. Little, R. Morandotti, A. Mitchell, and D. J. Moss, "Reconfigurable broadband microwave photonic intensity differentiator based on an integrated optical frequency comb source," *APL Photonics*, vol. 2, no. 9, 096104, Sep. 2017.

158. X. Xu, *et al.*, "Continuously tunable orthogonally polarized RF optical single sideband generator based on micro-ring resonators," *Journal of Optics,* vol. 20, no. 11, 115701. 2018.

159. X. Xu, *et al.*, "Orthogonally polarized RF optical single sideband generation and dual-channel equalization based on an integrated microring resonator," *Journal of Lightwave Technology,* vol. 36, no. 20, pp. 4808-4818. 2018.

160. X. Xu, *et al.,* "Photonic RF phase-encoded signal generation with a microcomb source", *J. Lightwave Technology*, vol. 38, no. 7, 1722-1727, 2020.

161. X. Xu et al, "Photonic perceptron based on a Kerr microcomb for scalable high speed optical neural networks", Laser and Photonics Reviews, vol. 14, no. 8, 2000070 (2020). DOI: 10.1002/lpor.202000070.

162. X. Xu, et al., "11 TOPs photonic convolutional accelerator for optical neural networks", Nature **589**, 44-51 (2021).

163. Xingyuan Xu, Weiwei Han, Mengxi Tan, Yang Sun, Yang Li, Jiayang Wu, Roberto Morandotti, Fellow IEEE, Arnan Mitchell, Senior Member IEEE, Kun Xu, and David J. Moss, Fellow IEEE Mengxi Tan, Xingyuan Xu, Jiayang Wu, Roberto Morandotti, Arnan Mitchell, and David J. Moss, "Neuromorphic computing based on wavelength-division multiplexing", **28** Early Access IEEE Journal of Selected Topics in Quantum Electronics Special Issue on Optical Computing. DOI:10.1109/JSTQE.2022.3203159.

164. Yang Sun, Jiayang Wu, Mengxi Tan, Xingyuan Xu, Yang Li, Roberto Morandotti, Arnan Mitchell, and David Moss, "Applications of optical micro-combs", Advances in Optics and Photonics (2023).

165. Yunping Bai, Xingyuan Xu,1, Mengxi Tan, Yang Sun, Yang Li, Jiayang Wu, Roberto Morandotti, Arnan Mitchell, Kun Xu, and David J. Moss, "Photonic multiplexing techniques for neuromorphic computing", accepted with modest revisions, Nanophotonics **12** (2023). DOI:10.1515/nanoph-2022-0485.

166. Chawaphon Prayoonyong, Andreas Boes, Xingyuan Xu, Mengxi Tan, Sai T. Chu, Brent E. Little, Roberto Morandotti, Arnan Mitchell, David J. Moss, and Bill Corcoran, "Frequency comb distillation for optical superchannel transmission", Journal of Lightwave Technology **39** (23) 7383-7392 (2021). DOI: 10.1109/JLT.2021.3116614.

167. Mengxi Tan, Xingyuan Xu, Jiayang Wu, Bill Corcoran, Andreas Boes, Thach G. Nguyen, Sai T. Chu, Brent E. Little, Roberto Morandotti, Arnan Mitchell, and David J. Moss, "Integral order photonic RF signal processors based on a soliton crystal micro-comb source", IOP Journal of Optics **23** (11) 125701 (2021). https://doi.org/10.1088/2040-8986/ac2eab

168. Kues, M. et al. "Quantum optical microcombs", Nature Photonics **13**, (3) 170-179 (2019). doi:10.1038/s41566-019-0363-0

169. C.Reimer, L. Caspani, M. Clerici, et al., "Integrated frequency comb source of heralded single photons," Optics Express, vol. 22, no. 6, pp. 6535-6546, 2014.

170. C.Reimer, et al., "Cross-polarized photon-pair generation and bi-chromatically pumped optical parametric oscillation on a chip", Nature Communications, vol. 6, Article 8236, 2015. DOI: 10.1038/ncomms9236.

171. L. Caspani, C. Reimer, M. Kues, et al., "Multifrequency sources of quantum correlated photon pairs on-chip: a path toward integrated Quantum Frequency Combs," Nanophotonics, vol. 5, no. 2, pp. 351-362, 2016.

172. P. Roztocki et al., "Practical system for the generation of pulsed quantum frequency combs," Optics Express, vol. 25, no. 16, pp. 18940-18949, 2017.

173. Y. Zhang, et al., "Induced photon correlations through superposition of two four-wave mixing processes in integrated cavities", Laser and Photonics Reviews, vol. 14, no. 7, pp. 2000128, 2020. DOI: 10.1002/lpor.202000128

174. C. Reimer, et al., "High-dimensional one-way quantum processing implemented on d-level cluster states", Nature Physics, vol. 15, no.2, pp. 148–153, 2019.

175. P.Roztocki et al., "Complex quantum state generation and coherent control based on integrated frequency combs", Journal of Lightwave Technology **37** (2) 338-347 (2019).

176. S. Sciara et al., "Generation and Processing of Complex Photon States with Quantum Frequency Combs", IEEE Photonics Technology Letters **31** (23) 1862-1865 (2019). DOI: 10.1109/LPT.2019.2944564.

177. Stefania Sciara, Piotr Roztocki, Bennet Fisher, Christian Reimer, Luis Romero Cortez, William J. Munro, David J. Moss, Alfonso C. Cino, Lucia Caspani, Michael Kues, J. Azana, and Roberto Morandotti, "Scalable and effective multilevel entangled photon states: A promising tool to boost quantum technologies", Nanophotonics 10 (18), 4447–4465 (2021). DOI:10.1515/nanoph-2021-0510.

178. L. Caspani, C. Reimer, M. Kues, et al., "Multifrequency sources of quantum correlated photon pairs on-chip: a path toward integrated Quantum Frequency Combs," Nanophotonics, vol. 5, no. 2, pp. 351-362, 2016.

179. Yuning Zhang, Jiayang Wu, Yang Qu, Yunyi Yang, Linnan Jia, Baohua Jia, and David J. Moss, "Enhanced supercontinuum generated in SiN waveguides coated with GO films", Advanced Materials Technologies **8** (2023). DOI: 10.1002/admt.202201796.

180. Yuning Zhang, Jiayang Wu, Linnan Jia, Yang Qu, Baohua Jia, and David J. Moss, "Graphene oxide for nonlinear integrated photonics", Laser and Photonics Reviews **17** (2023). DOI:10.1002/lpor.202200512.

181. Jiayang Wu, H.Lin, D. J. Moss, T.K. Loh, Baohua Jia, "Graphene oxide: new opportunities for electronics, photonics, and optoelectronics", Nature Reviews Chemistry **7** (2023). DOI:10.1038/s41570-022-00458-7.

182. Yang Qu, Jiayang Wu, Yuning Zhang, Yunyi Yang, Linnan Jia, Baohua Jia, and David J. Moss, "Photo thermal tuning in GO-coated integrated waveguides", Micromachines **13** 1194 (2022). doi.org/10.3390/mi13081194

183. Yuning Zhang, Jiayang Wu, Yunyi Yang, Yang Qu, Houssein El Dirani, Romain Crochemore, Corrado Sciancalepore, Pierre Demongodin, Christian Grillet, Christelle Monat, Baohua Jia, and David J. Moss, "Enhanced self-phase modulation in silicon nitride waveguides integrated with 2D graphene oxide films", IEEE Journal of Selected Topics in Quantum Electronics **28** Early Access (2022). DOI: 10.1109/JSTQE.2022.3177385

184. Yuning Zhang, Jiayang Wu, Yunyi Yang, Yang Qu, Linnan Jia, Baohua Jia, and David J. Moss, "Enhanced spectral broadening of femtosecond optical pulses in silicon nanowires integrated with 2D graphene oxide films", Micromachines **13** 756 (2022). DOI:10.3390/mi13050756.

185. Linnan Jia, Jiayang Wu, Yuning Zhang, Yang Qu, Baohua Jia, Zhigang Chen, and David J. Moss, "Fabrication Technologies for the On-Chip Integration of 2D Materials", Small: Methods **6**, 2101435 (2022). DOI:10.1002/smtd.202101435.

186. Yuning Zhang, Jiayang Wu, Yang Qu, Linnan Jia, Baohua Jia, and David J. Moss, "Design and optimization of four-wave mixing in microring resonators integrated with 2D graphene oxide films", Journal of Lightwave Technology **39** (20) 6553-6562 (2021). DOI:10.1109/JLT.2021.3101292. Print ISSN: 0733-8724, Online ISSN: 1558-2213 (2021).

187. Yuning Zhang, Jiayang Wu, Yang Qu, Linnan Jia, Baohua Jia, and David J. Moss, "Optimizing the Kerr nonlinear optical performance of silicon waveguides integrated with 2D graphene oxide films", Journal of Lightwave Technology 39 (14) 4671-4683 (2021). DOI: 10.1109/JLT.2021.3069733.

188. Jiayang Wu, Linnan Jia, Yuning Zhang, Yang Qu, Baohua Jia, and David J. Moss," Graphene oxide: versatile films for flat optics to nonlinear photonic chips", Advanced Materials **33** (3) 2006415, pp.1-29 (2021). DOI:10.1002/adma.202006415.

189. Y. Qu, J. Wu, Y. Zhang, L. Jia, Y. Yang, X. Xu, S. T. Chu, B. E. Little, R. Morandotti, B. Jia, and D. J. Moss, "Graphene oxide for enhanced optical nonlinear performance in CMOS compatible integrated devices", Paper No. 11688-30, PW21O-OE109-36, 2D Photonic Materials and Devices IV, SPIE Photonics West, San Francisco CA March 6-11 (2021). doi.org/10.1117/12.2583978

190. Jiayang Wu, Yunyi Yang, Yang Qu, Xingyuan Xu, Yao Liang, Sai T. Chu, Brent E. Little, Roberto Morandotti, Baohua Jia, and David J. Moss, "Graphene oxide waveguide polarizers and polarization selective micro-ring resonators", Paper 11282-29, SPIE Photonics West, San Francisco, CA, 4 - 7 February (2020). doi: 10.1117/12.2544584

191. Jiayang Wu, Yunyi Yang, Yang Qu, Xingyuan Xu, Yao Liang, Sai T. Chu, Brent E. Little, Roberto Morandotti, Baohua Jia, and David J. Moss, "Graphene oxide waveguide polarizers and polarization selective micro-ring resonators", Laser and Photonics Reviews vol. **13** (9) 1900056 (2019). DOI:10.1002/lpor.201900056.

192. A. Pasquazi, et al., "Sub-picosecond phase-sensitive optical pulse characterization on a chip", Nature Photonics, vol. 5, no. 10, pp. 618-623 (2011).

193. Bao, C., et al., Direct soliton generation in microresonators, Opt. Lett, **42**, 2519 (2017).

194. M. Kues, et al., "Passively modelocked laser with an ultra-narrow spectral width", Nature Photonics, vol. 11, no. 3, pp. 159, 2017.

195. L. Razzari, et al., "CMOS-compatible integrated optical hyper-parametric oscillator," Nature Photonics, vol. 4, no. 1, pp. 41-45, 2010.

196. M. Ferrera, et al., "Low-power continuous-wave nonlinear optics in doped silica glass integrated waveguide structures," Nature Photonics, vol. 2, no. 12, pp. 737-740, 2008.

197. M.Ferrera et al."On-Chip ultra-fast 1st and 2nd order CMOS compatible all-optical integration", Opt. Express, vol. 19, (23)pp. 23153-23161 (2011).

198. D. Duchesne, M. Peccianti, M. R. E. Lamont, et al., "Supercontinuum generation in a high index doped silica glass spiral waveguide," Optics Express, vol. 18, no, 2, pp. 923-930, 2010.

199. H Bao, L Olivieri, M Rowley, ST Chu, BE Little, R Morandotti, DJ Moss, ... "Turing patterns in a fiber laser with a nested microresonator: Robust and controllable microcomb generation", Physical Review Research **2** (2), 023395 (2020).

200. M. Ferrera, et al., "On-chip CMOS-compatible all-optical integrator", Nature Communications, vol. 1, Article 29, 2010.

201. A. Pasquazi, et al., "All-optical wavelength conversion in an integrated ring resonator," Optics Express, vol. 18, no. 4, pp. 3858-3863, 2010.

202. A.Pasquazi, Y. Park, J. Azana, et al., "Efficient wavelength conversion and net parametric gain via Four Wave Mixing in a high index doped silica waveguide," Optics Express, vol. 18, no. 8, pp. 7634-7641, 2010.

203. M. Peccianti, M. Ferrera, L. Razzari, et al., "Subpicosecond optical pulse compression via an integrated nonlinear chirper," Optics Express, vol. 18, no. 8, pp. 7625-7633, 2010.

204. Little, B. E. et al., "Very high-order microring resonator filters for WDM applications", IEEE Photonics Technol. Lett. **16**, 2263–2265 (2004).

205. M. Ferrera et al., "Low Power CW Parametric Mixing in a Low Dispersion High Index Doped Silica Glass Micro-Ring Resonator with Q-factor > 1 Million", Optics Express, vol.17, no. 16, pp. 14098–14103 (2009).

206. M. Peccianti, et al., "Demonstration of an ultrafast nonlinear microcavity modelocked laser", Nature Communications, vol. 3, pp. 765, 2012.
207. A.Pasquazi, et al., "Self-locked optical parametric oscillation in a CMOS compatible microring resonator: a route to robust optical frequency comb generation on a chip," Optics Express, vol. 21, no. 11, pp. 13333-13341, 2013.
208. A.Pasquazi, et al., "Stable, dual mode, high repetition rate mode-locked laser based on a microring resonator," Optics Express, vol. 20, no. 24, pp. 27355-27362, 2012.
209. Pasquazi, A. et al. Micro-combs: a novel generation of optical sources. Physics Reports **729**, 1-81 (2018).
210. Moss, D. J. et al., "New CMOS-compatible platforms based on silicon nitride and Hydex for nonlinear optics", Nature photonics **7**, 597 (2013).
211. H. Bao, et al., Laser cavity-soliton microcombs, Nature Photonics, vol. 13, no. 6, pp. 384-389, Jun. 2019.
212. Antonio Cutrona, Maxwell Rowley, Debayan Das, Luana Olivieri, Luke Peters, Sai T. Chu, Brent L. Little, Roberto Morandotti, David J. Moss, Juan Sebastian Totero Gongora, Marco Peccianti, Alessia Pasquazi, "High Conversion Efficiency in Laser Cavity-Soliton Microcombs", Optics Express Vol. 30, Issue 22, pp. 39816-39825 (2022). https://doi.org/10.1364/OE.470376.
213. M.Rowley, P.Hanzard, A.Cutrona, H.Bao, S.Chu, B.Little, R.Morandotti, D. J. Moss, G. Oppo, J. Gongora, M. Peccianti and A. Pasquazi, "Self-emergence of robust solitons in a micro-cavity", Nature **608** (7922) 303–309 (2022).

214. Hamed Arianfard, Saulius Juodkazis, David J. Moss, and Jiayang Wu, "Sagnac interference in integrated photonics", Applied Physics Reviews vol. **10** (2023).
215. Hamed Arianfard, Jiayang Wu, Saulius Juodkazis, and David J. Moss, "Spectral shaping based on optical waveguides with advanced Sagnac loop reflectors", Paper No. PW22O-OE201-20, SPIE-Opto, Integrated Optics: Devices, Materials, and Technologies XXVI, SPIE Photonics West, San Francisco CA January 22 - 27 (2022). doi: 10.1117/12.2607902
216. Hamed Arianfard, Jiayang Wu, Saulius Juodkazis, David J. Moss, "Spectral Shaping Based on Integrated Coupled Sagnac Loop Reflectors Formed by a Self-Coupled Wire Waveguide", IEEE Photonics Technology Letters vol. **33** (13) 680-683 (2021). DOI:10.1109/LPT.2021.3088089.
217. Hamed Arianfard, Jiayang Wu, Saulius Juodkazis and David J. Moss, "Three Waveguide Coupled Sagnac Loop Reflectors for Advanced Spectral Engineering", Journal of Lightwave Technology vol. **39** (11) 3478-3487 (2021). DOI: 10.1109/JLT.2021.3066256.
218. Hamed Arianfard, Jiayang Wu, Saulius Juodkazis and David J. Moss, "Advanced Multi-Functional Integrated Photonic Filters based on Coupled Sagnac Loop Reflectors", Journal of Lightwave Technology vol. **39** Issue: 5, pp.1400-1408 (2021). DOI:10.1109/JLT.2020.3037559.
219. Hamed Arianfard, Jiayang Wu, Saulius Juodkazis and David J. Moss, "Advanced multi-functional integrated photonic filters based on coupled Sagnac loop reflectors", Paper 11691-4, PW21O-OE203-44, Silicon Photonics XVI, SPIE Photonics West, San Francisco CA March 6-11 (2021). doi.org/10.1117/12.2584020
220. Jiayang Wu, Tania Moein, Xingyuan Xu, and David J. Moss, "Advanced photonic filters via cascaded Sagnac loop reflector resonators in silicon-on-insulator integrated nanowires", Applied Physics Letters Photonics vol. **3** 046102 (2018). DOI:/10.1063/1.5025833
221. Jiayang Wu, Tania Moein, Xingyuan Xu, Guanghui Ren, Arnan Mitchell, and David J. Moss, "Micro-ring resonator quality factor enhancement via an integrated Fabry-Perot cavity", Applied Physics Letters Photonics vol. **2** 056103 (2017). doi: 10.1063/1.4981392.

222. E.D Ghahramani, DJ Moss, JE Sipe, "Full-band-structure calculation of first-, second-, and third-harmonic optical response coefficients of ZnSe, ZnTe, and CdTe", Physical Review **B 43** (12), 9700 (1991).

223. C Grillet, C Smith, D Freeman, S Madden, B Luther-Davies, EC Magi, ... "Efficient coupling to chalcogenide glass photonic crystal waveguides via silica optical fiber nanowires", Optics Express vol. **14** (3), 1070-1078 (2006).
224. S Tomljenovic-Hanic, MJ Steel, CM de Sterke, DJ Moss, "High-Q cavities in photosensitive photonic crystals" Optics Letters vol. **32** (5), 542-544 (2007).
225. M Ferrera et al., "On-Chip ultra-fast 1st and 2nd order CMOS compatible all-optical integration", Optics Express vol. **19** (23), 23153-23161 (2011).

226. VG Ta'eed et al., "Error free all optical wavelength conversion in highly nonlinear As-Se chalcogenide glass fiber", Optics Express vol. **14** (22), 10371-10376 (2006).
227. M Rochette, L Fu, V Ta'eed, DJ Moss, BJ Eggleton, "2R optical regeneration: an all-optical solution for BER improvement", IEEE Journal of Selected Topics in Quantum Electronics vol. **12** (4), 736-744 (2006).
228. TD Vo, et al., "Silicon-chip-based real-time dispersion monitoring for 640 Gbit/s DPSK signals", Journal of Lightwave Technology vol. **29** (12), 1790-1796 (2011).